\documentclass{PoS}

\title{Hard and Soft Physics at RHIC \\ with implications for LHC}

\ShortTitle{Hard and Soft Physics at RHIC with implications for LHC}

\author{\speaker{Michael J.  Tannenbaum}%
         \thanks{Supported by the U.S. Department of Energy, Contract No. DE-AC02-98CH1-886.}\\
         Physics Dept., 510c, Brookhaven National Laboratory, Upton, NY 11973-5000,USA\\
        E-mail: \email{mjt@bnl.gov}}

\abstract{Measurements of $\pi^0$, identified charged hadrons, direct-single-$\gamma$, direct single-$e^{\pm}$  and jets are expected to be important probes of the Quark Gluon Plasma (QGP) at LHC as they are at RHIC. Recent results from RHIC will be presented, with implications, possibly grave, for LHC. Issues of soft-physics, such as $E_T$  distributions, will be discussed with questions as to: whether the soft multipicity or sum-transverse energy ($E_T$), may be the only quantity at LHC to exhibit number of collision scaling? or whether the anisotropic flow ($v_2$) will exceed the hydrodynamic limit?

 }
\FullConference{4th international workshop High-pT physics at LHC 09\\
		 February 4-7, 2009\\
		 Prague, Czech Republic}
\usepackage{calc}
\newcounter{hours}\newcounter{minutes}

\usepackage{epsfig} 
\def\lsim{\raise0.3ex\hbox{$<$\kern-0.75em\raise-1.1ex\hbox{$\sim$}}}
\def\gsim{\raise0.3ex\hbox{$>$\kern-0.75em\raise-1.1ex\hbox{$\sim$}}}

\def\mean#1{\left<#1\right>}
\def\Journal#1#2#3#4{ {\it{#1}} {\bf #2}, #3 (#4)}
\def\IJMPA{{Int. J. Mod. Phys.}~{\rm A}}

\def\EPJC{{Eur. Phys. J.}~{\rm C}}
\def\EPJST{{Eur. Phys. J.}~{Special Topics}\ }

\def\JPG{{J. Phys.}~{\rm G}}

\def\NPA{{Nucl. Phys.}~{\rm A}}
\def\NPB{{Nucl. Phys.}~{\rm B}}
\def\PLB{{Phys. Lett.}~{\rm B}}

\def\PR{Phys. Rev.\ } 
\def\PRL{Phys. Rev. Lett.\ }
\def\PRD{{Phys. Rev.}~{\rm D}}
\def\PRC{{Phys. Rev.}~{\rm C}}

\def\ZPC{{Z. Phys.}~{\rm C}}
\def\ARNPS{{Ann. Rev. Nucl. Part. Sci.\ }}

\begin{document}

\section{Large Transverse Momentum $\pi^0$ production---from ISR to RHIC}
\begin{figure}[!thb]
\begin{center}
\begin{tabular}{cc}
\includegraphics[width=0.50\linewidth]{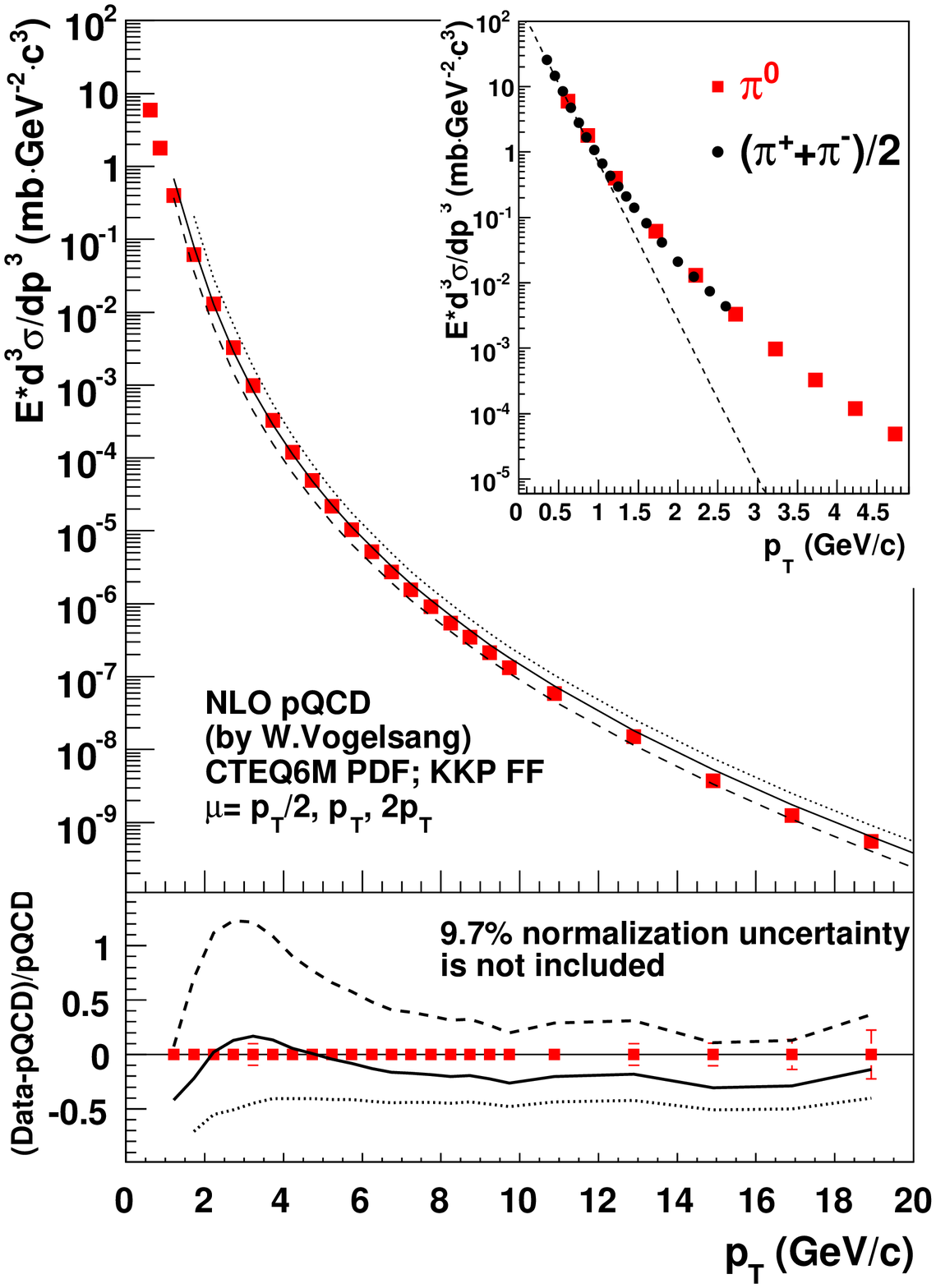}&
\hspace*{-0.02\linewidth}\includegraphics[width=0.56\linewidth]{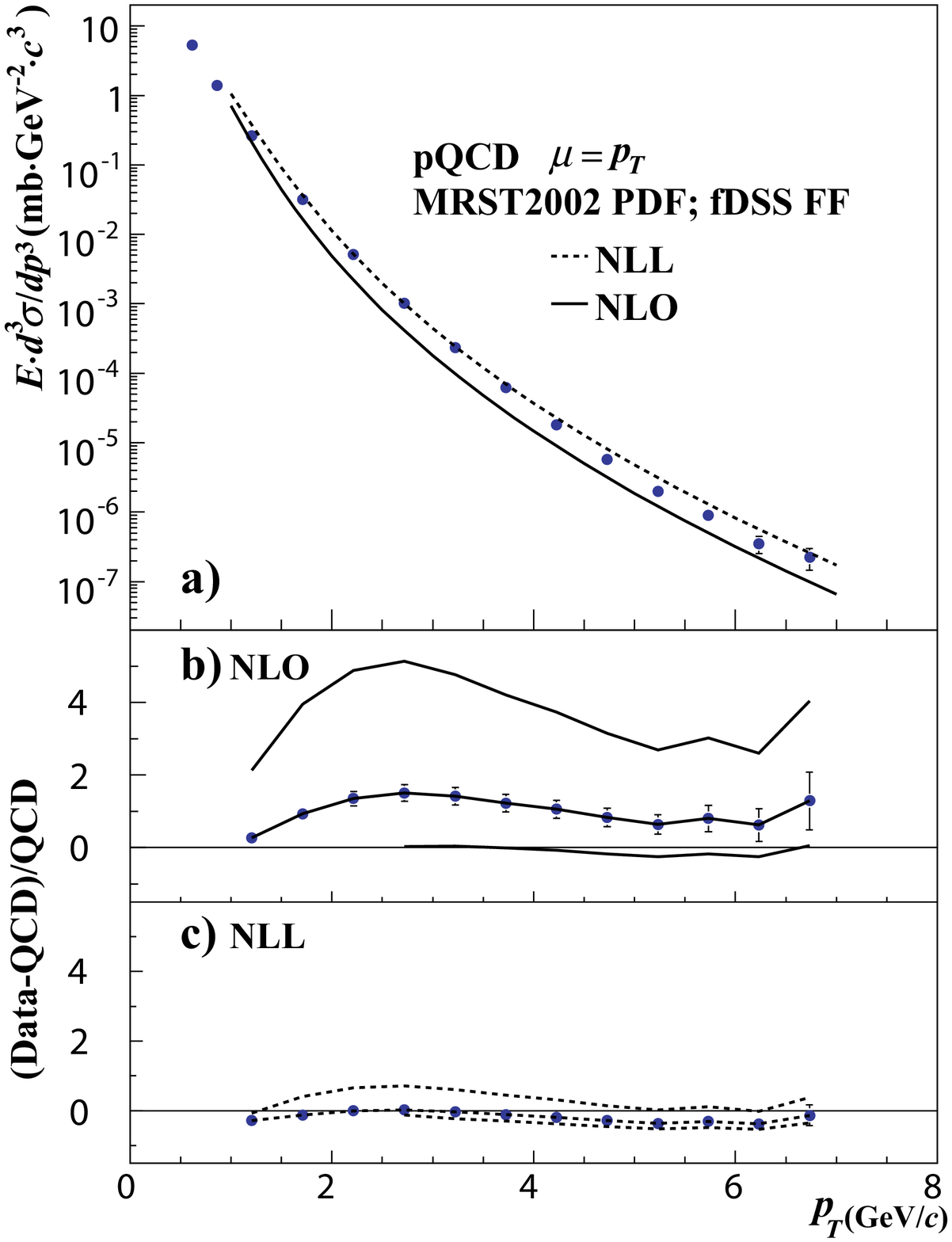}
\end{tabular}
\end{center}\vspace*{-0.25in}
\caption[]{(left) PHENIX measurement of invariant cross section, $E {d^3\sigma}/{d^3p}$, as a function of transverse momentum $p_T$ for $\pi^0$ production at mid-rapidity in p-p collisions at c.m. energy $\sqrt{s}=200$~GeV~\cite{ppg063}. (right) PHENIX measurement of $\pi^0$ in p-p collisions at $\sqrt{s}=62.4$~GeV~\cite{ppg087}.  }
\label{fig:PXpi0pp}
\end{figure}

PHENIX has presented measurements of $\pi^0$ production at mid-rapidity in p-p collisions at two values of c.m. energy $\sqrt{s}$=200~GeV and 62.4~GeV (Fig.~\ref{fig:PXpi0pp}). Some of my younger colleagues are amazed at the excellent agreement of Next to Leading Order (NLO) and Next to Leading Log (NLL) perturbative Quantum Chromodynamics (pQCD) calculations~\cite{ppg063,ppg087} with the measurements. 
However, this comes as no surprise to me because hard scattering in p-p collisions was discovered at the CERN ISR by the observation of a very large flux of high transverse momentum $\pi^0$ with a power-law tail which varied systematically with the c.m. energy of the collision. This observation in 1972 proved that the partons of deeply inelastic scattering were strongly interacting. Further ISR measurements utilizing inclusive single or pairs of hadrons established that high transverse momentum particles are produced from states with two roughly back-to-back jets which are the result of scattering of point-like constituents of the protons as described by QCD, which was developed during the course of these measurements.  
  \subsection{ISR Data, Notably CCR 1972-73} 
\begin{figure}[ht]
\begin{center}
\begin{tabular}{cc}
\psfig{file=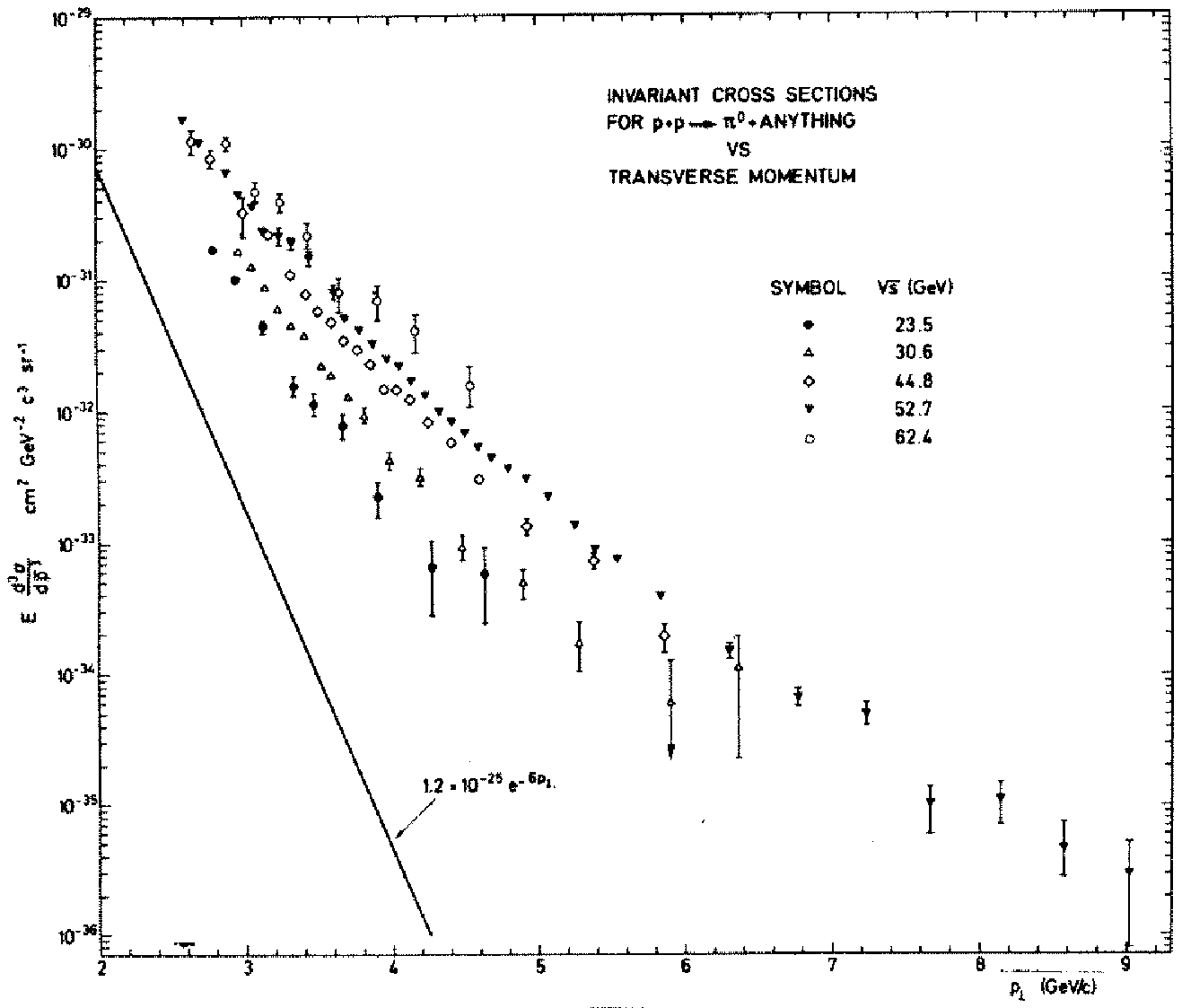,width=0.48\linewidth}\hspace*{0.014in}
\psfig{file=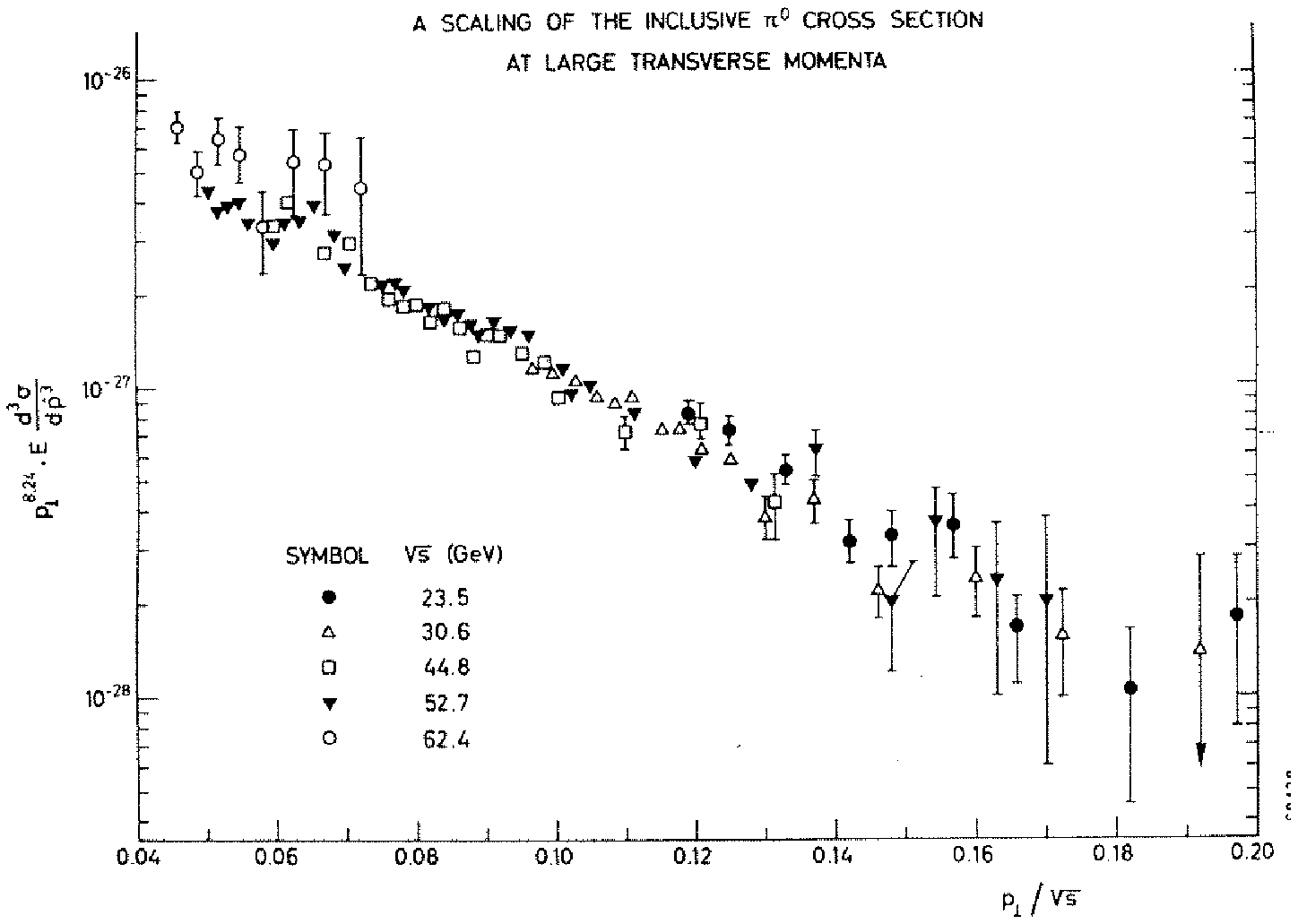,width=0.50\linewidth}
\end{tabular}
\end{center}
\caption[]
{(left) CCR~\cite{CCR} transverse momentum dependence of the invariant cross section at five center of mass energies. (right) The above data multiplied by $p_{T}^n$, using the best fit value of $n=8.24\pm0.05$, with $F=Ae^{-b x_{T}}$, plotted vs $p_{T}/\sqrt{s}$ $(=x_T/2)$.   
\label{fig:CCR} }
\end{figure}

The Cern Columbia Rockefeller (CCR) Collaboration~\cite{CCR} 
(and also the Saclay Strasbourg~\cite{SS} and British Scandinavian~\cite{BS} 
collaborations) measured 
high $p_T$ pion production at the CERN-ISR (Fig.~\ref{fig:CCR}). 
The $e^{-6p_T}$ breaks to a power law at high $p_T$ with 
characteristic $\sqrt{s}$ dependence.~\footnote{The clear break of the exponential to a power law at $\sqrt{s}=200$~GeV is shown in the inset of Fig.~\ref{fig:PXpi0pp}-(left).}  
The large rate indicates that {\em partons interact 
strongly ($\gg$ EM) with each other}, {\bf but,} to quote the authors~\cite{CCR}: 
``Indeed, the possibility of a break in the steep exponential slope observed 
at low $p_T$ was anticipated by Berman, Bjorken and Kogut~\cite{BBK}. However, the  electromagnetic form they predict, $p_{T}^{-4} F(p_{T}/\sqrt{s})$, is not observed in our experiment. On the other hand, a constituent exchange 
model proposed by Blankenbacler, Brodsky and Gunion~\cite{CIM}, and extended by others, 
does give an excellent account of the data.''   
The data fit 
$p_{T}^{-n} F(p_{T}/\sqrt{s})$, with $n\simeq8$. 

\subsection{Constituent Interchange Model (CIM) 1972}

Inspired by the {\em dramatic features} 
of pion inclusive reactions revealed by ``the recent measurements at CERN 
ISR of single-particle inclusive scattering at $90^\circ$ and large 
transverse momentum'', Blankenbecler, Brodsky and Gunion~\cite{CIM} 
proposed a new general scaling form: 
\begin{equation}
E \frac{d^3\sigma}{dp^3}=\frac{1}{p_T^{n}} F({p_T \over \sqrt{s}}) 
\label{eq:bbg}
\end{equation}
where $n$ gives the form of the force-law 
between constituents. For QED or Vector Gluon exchange, $n=4$, but 
perhaps more importantly, BBG predict 
{\bf $n$=8} for the case of quark-meson scattering by the exchange of 
a quark (CIM) as apparently observed. 

\subsection{First prediction using `QCD' 1975---WRONG!}

R.~F.~Cahalan, K.~A.~Geer, J.~Kogut and Leonard Susskind~\cite{CGKS} 
generalized, in their own words: ``The naive, pointlike parton 
model of Berman, Bjorken and Kogut to scale-invariant and 
asymptotically free field theories. The asymptotically free field 
generalization is studied in detail. Although such theories contain vector 
fields, {\bf{ single vector-gluon exchange contributes insignificantly to 
wide-angle hadronic collisions.}} This follows from (1) the smallness of the 
invariant charge at small distances and (2) the {\em breakdown of naive 
scaling} in these theories. These effects should explain the apparent absence 
of vector exchange in inclusive and exclusive hadronic collisions at 
large momentum transfers observed at Fermilab and at the CERN ISR.''\footnote{There is an acknowledgement in this paper which is worthy of note:``Two of us (J.~K. and L.~S.) also thank S.~Brodsky for {\em emphasizing to us \underline{repeatedly}} that the present data on wide-angle hadron scattering {\em show no evidence for vector exchange.''}}  
 
	Nobody's perfect, they get {\em one} thing right! 
They introduce the ``effective index'' $n_{\rm eff}(x_T, \sqrt{s})$ to account for 
`scale breaking':
\begin{equation}
E \frac{d^3\sigma}{dp^3}={1 \over {p_T^{{n_{\rm eff}(x_T,\sqrt{s})}} }  } 
F({x_T})={1\over {\sqrt{s}^{{\,n_{\rm eff}(x_T,\sqrt{s})}} } } 
\: G({x_T})\qquad , 
\label{eq:nxt}
\end{equation}
where $x_T=2p_T/\sqrt{s}$.
\subsection{CCOR 1978---Higher $p_T>7$~GeV/c---$n_{\rm eff}(x_T, \sqrt{s}) \rightarrow 5=4^{++}$. QCD works! } 

\begin{figure}[hbt]
\begin{center}
\begin{tabular}{cc}
\begin{tabular}[b]{c}
\psfig{file=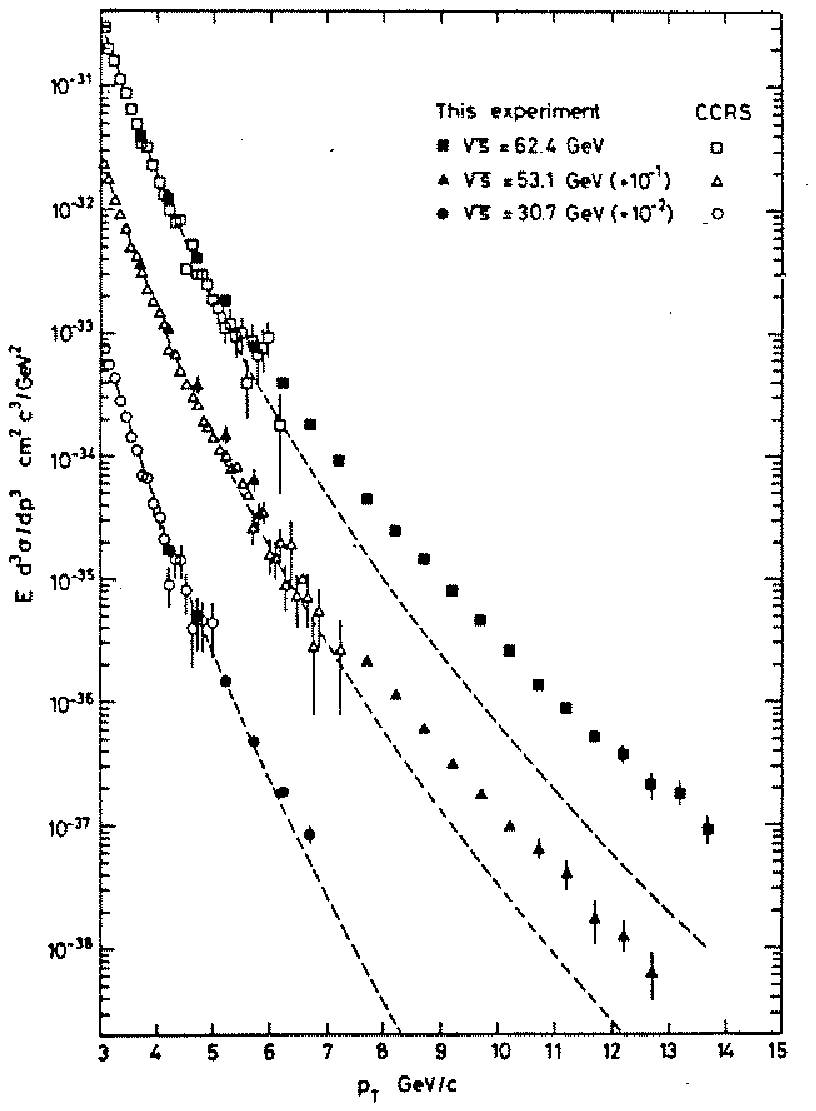,width=0.40\linewidth}\hspace*{0.014in}\cr
\begin{picture}(50,30)
\end{picture}
\end{tabular}
\begin{tabular}[b]{c}
\psfig{file=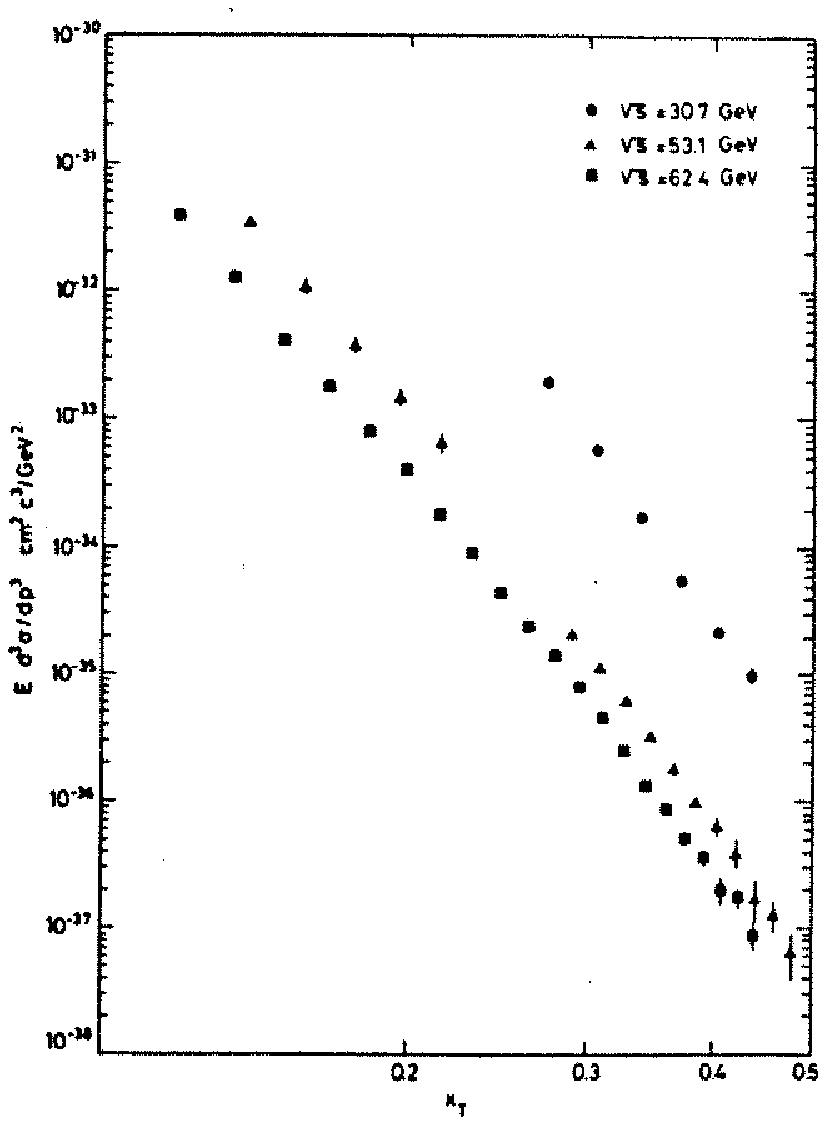,width=0.30\linewidth}\cr
\vspace*{0.15in}
\hspace*{0.1in}\psfig{file=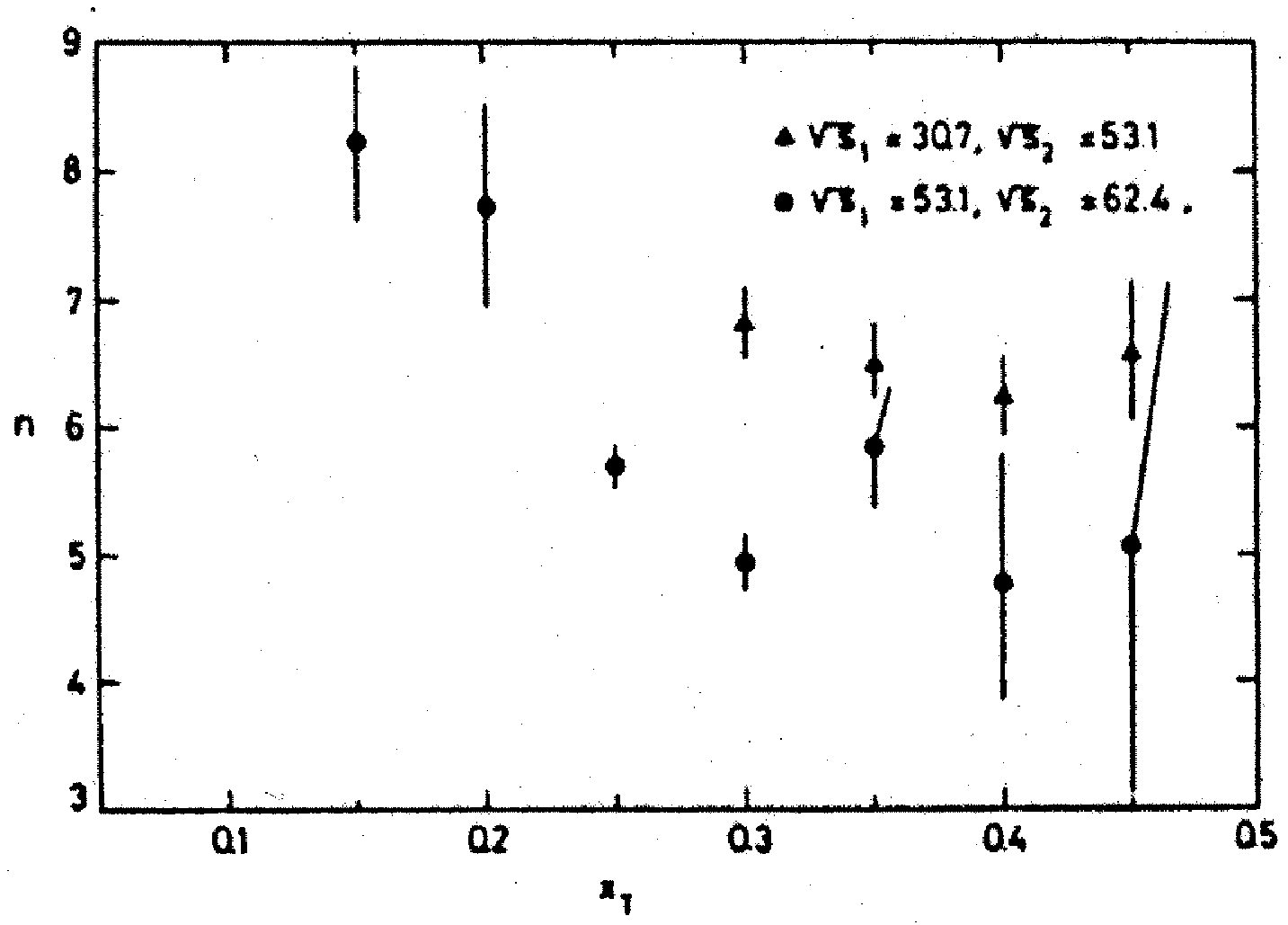,width=0.30\linewidth}
\end{tabular}
\end{tabular}

\end{center}
\vspace*{-0.28in}
\caption[]
{a) (left) CCOR~\cite{CCOR} transverse momentum dependence of the invariant cross section for 
\mbox{$p+p\rightarrow \pi^0 + X$} at 
three center of mass energies. Cross sections are offset by the factors 
noted. Open points and dashed fit are from a previous experiment, 
CCRS~\cite{CCRS}. b) (right)-(top) Same CCOR invariant cross sections plotted vs $x_T=2 p_T/\sqrt{s}$ on a log-log scale. c) (right)-(bottom) $n_{\rm eff}(x_T,\sqrt{s})$ derived from the combinations indicated.  The systematic normalization error at $\sqrt{s}=30.6$~GeV has been added in quadrature. There is an additional common systematic error of $\pm 0.33$ in $n$.  
\label{fig:ccorpt} }
\end{figure}

The CCOR measurement~\cite{CCOR} (Fig.~\ref{fig:ccorpt}) with a larger 
apparatus and much increased integrated 
luminosity extended their previous $\pi^0$ measurement~\cite{CCR,CCRS} to 
much higher $p_T$. 
The $p_T^{-8}$ scaling-fit  which worked at lower $p_T$
extrapolated below the higher $p_T$ measurements for $\sqrt{s} > 30.7$~GeV and 
$p_T \geq 7$~GeV/c (Fig.~\ref{fig:ccorpt}a). A fit to the new data~\cite{CCOR} for $7.5\leq p_T\leq 14.0$~GeV/c, $53.1\leq \sqrt{s}\leq 62.4$~GeV gave    
\mbox{$E d^3 \sigma/dp^3\simeq p_T^{-{5.1\pm 0.4}} (1-x_T)^{12.1\pm 0.6}$}, 
(including {\em all} systematic errors). 

	The effective index $n_{\rm eff}(x_T, \sqrt{s})$ was also extracted point-by-point from the data as shown in Fig.~\ref{fig:ccorpt}b where the CCOR data of Fig.~\ref{fig:ccorpt}a for 
the 3 values of $\sqrt{s}$ are plotted vs $x_T$ on a log-log scale. $n_{\rm eff}(x_T, \sqrt{s})$ is determined for any 2 values of $\sqrt{s}$ by taking 
the ratio as a function of $x_T$ as shown in 
Fig.~\ref{fig:ccorpt}c. $n_{\rm eff}(x_T, \sqrt{s})$  clearly varies 
with both $\sqrt{s}$ and $x_T$, it is not a constant. For 
$\sqrt{s}=53.1$ and 62.4~GeV, $n_{\rm eff}(x_T, \sqrt{s})$ varies from $\sim 8$ at low 
$x_T$ to $\sim 5$ at high $x_T$. An important feature of the scaling analysis (Eq.~\ref{eq:nxt})  
relevant to determining $n_{\rm eff}(x_T, \sqrt{s})$ is that {\em the absolute $p_T$ scale uncertainty cancels!}

\begin{figure}[ht]
\begin{center}
\begin{tabular}[b]{cc}
\begin{tabular}[b]{c}
\hspace*{-0.1in}\psfig{file=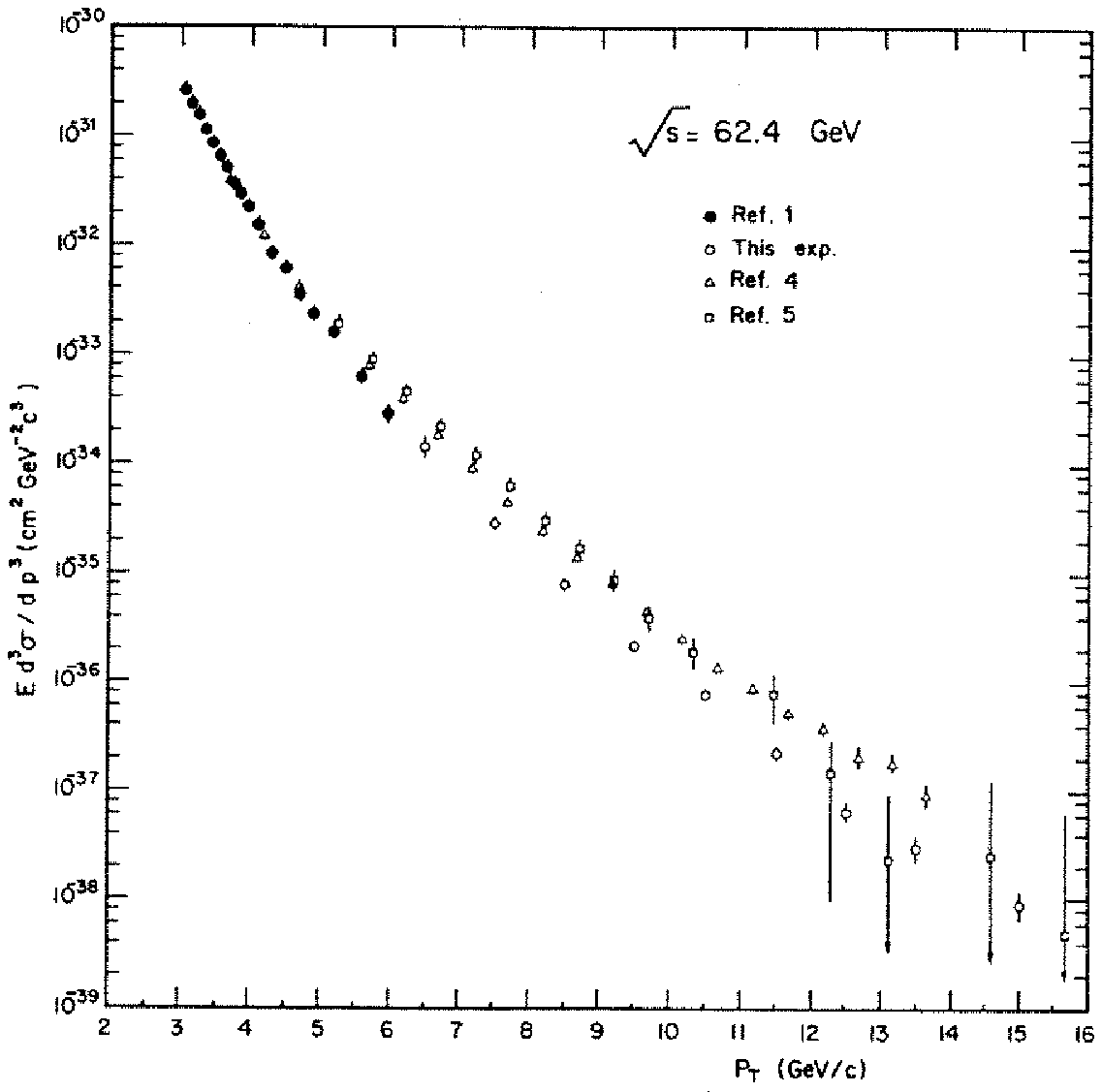,width=0.40\linewidth}\cr
\hspace*{-0.1in}\vspace*{0.15in}\psfig{file=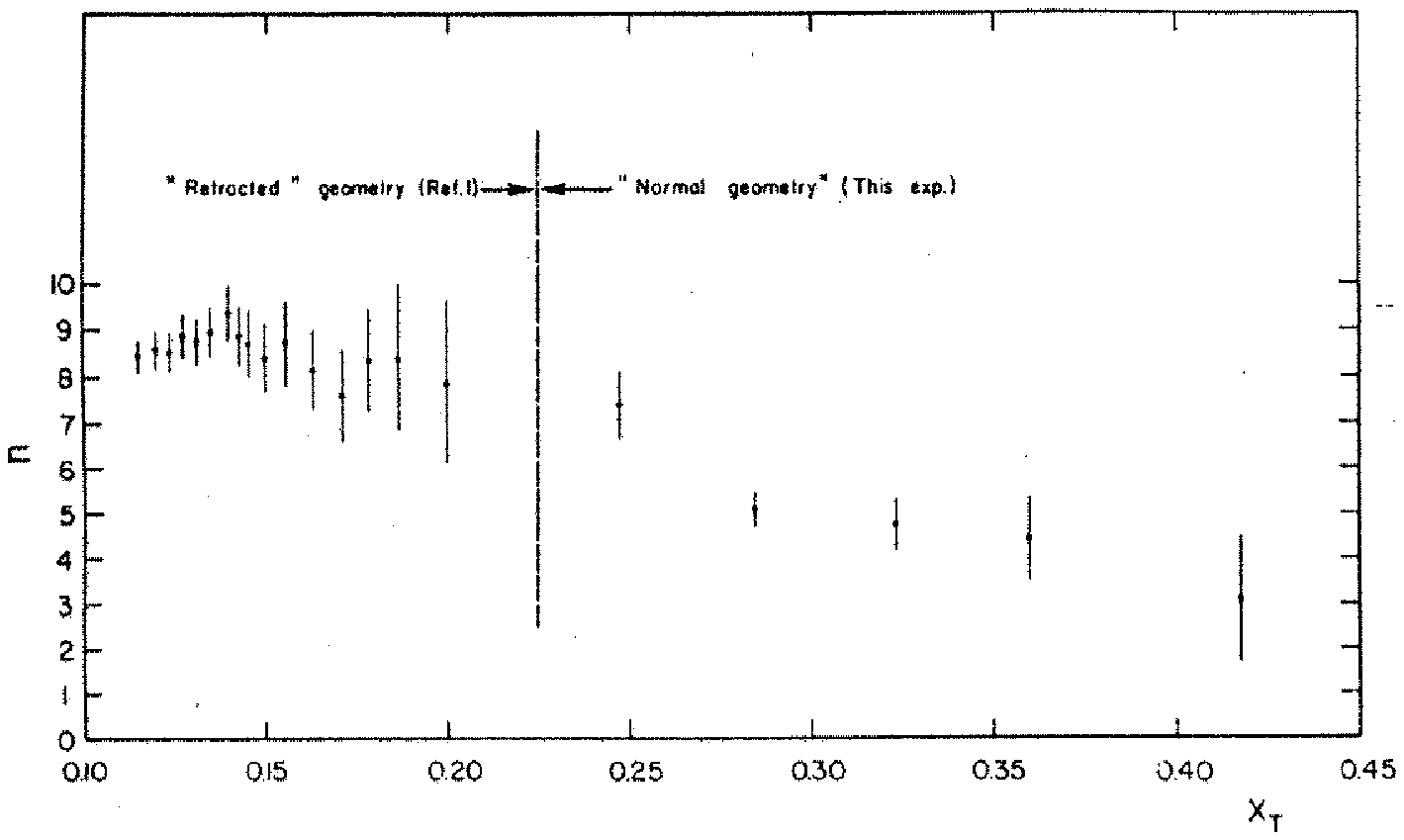,width=0.45\linewidth}
\end{tabular}
\begin{tabular}[b]{c}
\includegraphics[width=0.55\linewidth,height=0.68\linewidth]{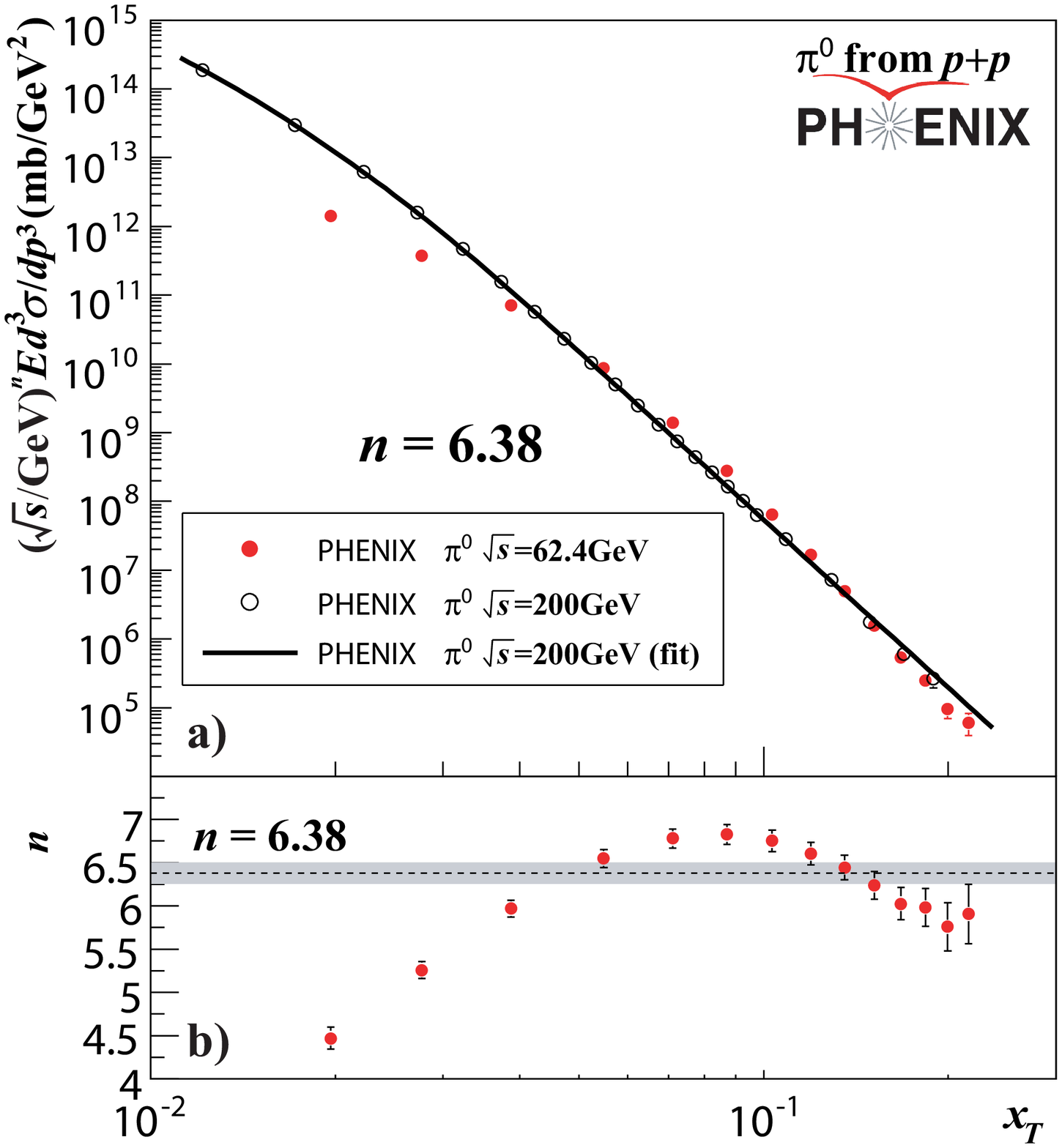}\cr
\begin{picture}(50,10)
\end{picture}
\end{tabular}
\end{tabular}
\end{center}
\vspace*{-0.18in}
\caption[]
{(left)-(top) Invariant cross section for inclusive $\pi^0$ for several ISR 
experiments, compiled by ABCS Collaboration~\cite{ABCS}; (left)-(bottom) $n_{\rm eff}(x_T,\sqrt{s})$ from ABCS 52.7, 62.4~GeV data only. There is an additional common systematic error of $\pm0.7$ in $n$. (right)-a) $\sqrt{s}{\rm (GeV)}^{6.38} \times Ed^3\sigma/dp^3$ as a function of $x_T=2p_T/\sqrt{s}$ for the PHENIX 62.4 and 200~GeV $\pi^0$ data from Fig.~\ref{fig:PXpi0pp}; (right)-b) point-by-point $n_{\rm eff}(x_T,\sqrt{s})$. 
\label{fig:otherxt} }
\end{figure}
The effect of the absoulte scale uncertainty, which 
is the main systematic error in these experiments,  can be gauged from 
Fig.~\ref{fig:otherxt}-(left)-(top)~\cite{ABCS} which shows the $\pi^0$ cross 
sections from several experiments. The absolute cross sections disagree by 
factors of $\sim 3$ for different experiments but the values of 
$n_{\rm eff}(x_T, \sqrt{s})$ for the CCOR~\cite{CCOR} 
(Fig.~\ref{fig:ccorpt}-(right)-(bottom)) and ABCS~\cite{ABCS} experiment 
(Fig.~\ref{fig:otherxt}-(left)-(bottom)) are in excellent agreement due 
to the cancellation of the error in the absolute $p_T$ scale. The $x_T$ scaling of the PHENIX p-p $\pi^0$ data at $\sqrt{s}=200$ and 62.4~GeV from Fig.~\ref{fig:PXpi0pp} with  
$n_{\rm eff}(x_T, \sqrt{s})\approx 6.38$ is shown in Fig.~\ref{fig:otherxt}-(right). The log-log plot emphasizes the pure power-law $p_T$ dependence of the invariant cross section, $E d^3\sigma/dp^3\simeq p_T^{-n}$ for $p_T >4$~GeV/c, with $n=8.11\pm 0.05$ at $\sqrt{s}=200$~GeV.~\cite{ppg087}  

	The first modern QCD calculations and 
predictions for high $p_T$ single particle inclusive cross sections, including 
non-scaling and initial state radiation were done in 1978, by Jeff 
Owens and collaborators.~\cite{Owens78} Jets in $4\pi$ Calorimeters at ISR 
energies or lower are invisible below $\sqrt{\hat{s}}\sim E_T \leq 25$ 
GeV~\cite{Gordon}; but there were many false claims which led to skepticism 
about jets in hadron collisions, particularly in the USA.~\cite{MJTIJMPA}  
A `phase change' in belief-in-Jets was produced by one UA2 event 
at the 1982 ICHEP in Paris~\cite{Paris82}, but that's another story.~\cite{MJTJyv1}  

\section{The major discovery at RHIC--$\pi^0$ suppression in A+A collisions.}
\begin{figure}[!ht]
\begin{center}
\begin{tabular}{cc}
\hspace*{-0.02\linewidth}\includegraphics[width=0.53\linewidth,height=0.4\linewidth]{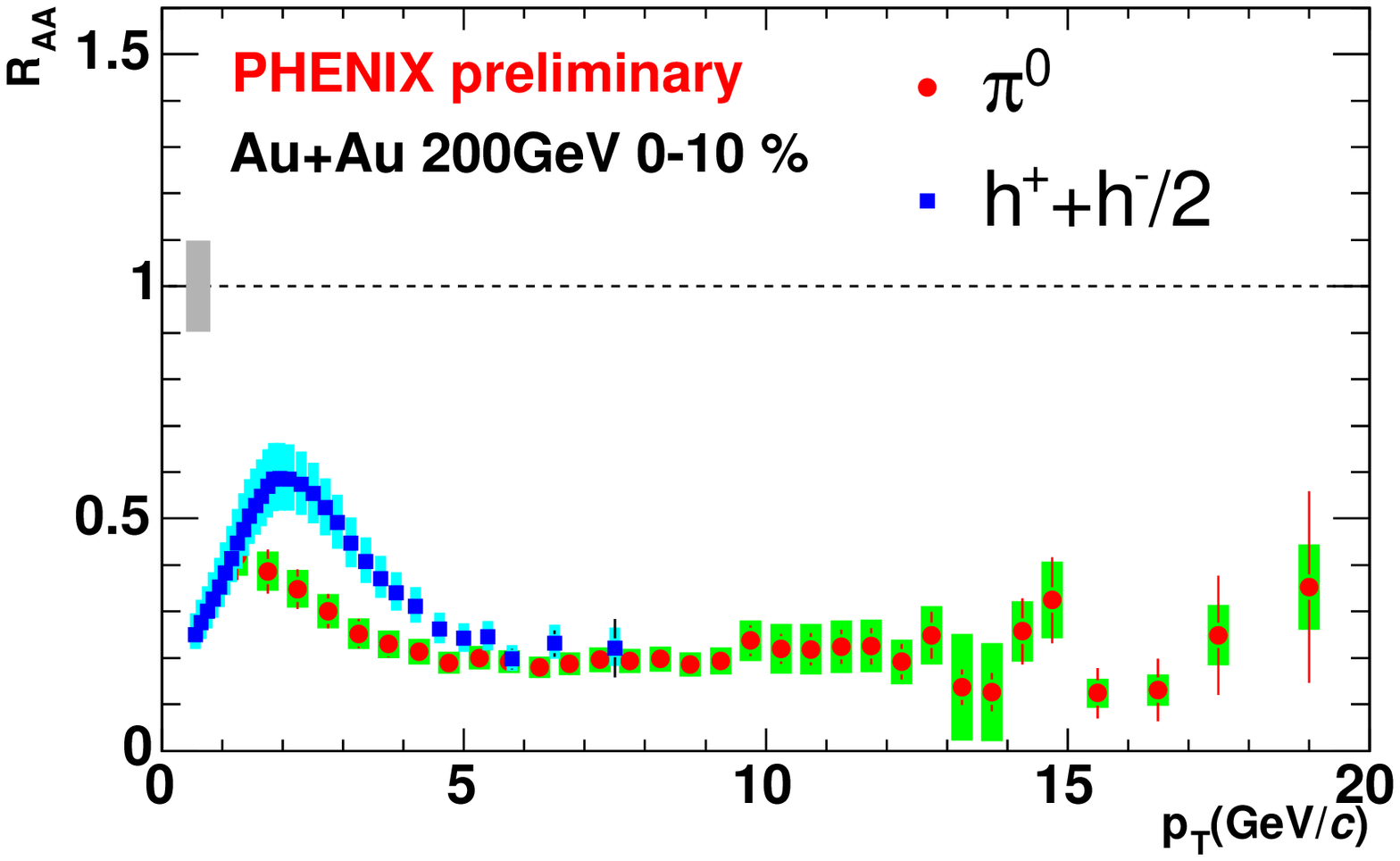} &\hspace*{-0.08\linewidth} 
\includegraphics[width=0.47\linewidth]{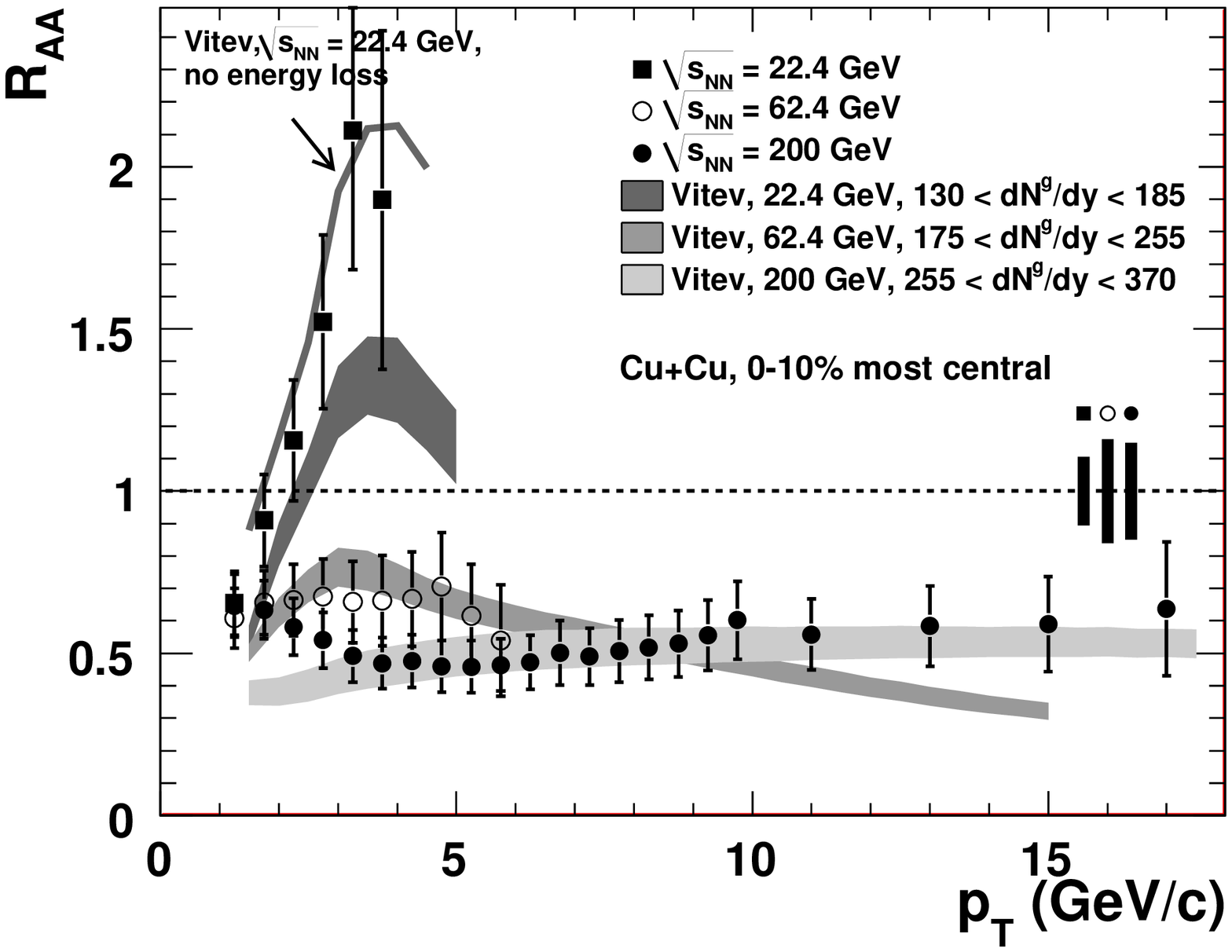} 
\end{tabular}
\end{center}
\caption[]{(left) Nuclear modification factor $R_{AA}(p_T)$ for $\pi^0$ and $h^{\pm}$ in central Au+Au collisions at $\sqrt{s_{NN}}=200$~GeV~\cite{pi0-QM05}; (right) $R_{AA}(p_T)$ for $\pi^0$ in Cu+Cu central collisions at $\sqrt{s_{NN}}=200$, 62.4 and 22.4~GeV~\cite{ppg084}, together with theory curves~\cite{Vitev2}. }
\label{fig:f2}
\end{figure}

   The discovery, at RHIC, that $\pi^0$ are suppressed by roughly a factor of 5 compared to point-like scaling of hard-scattering in central Au+Au collisions is arguably {\em the}  major discovery in Relativistic Heavy Ion Physics. In Fig.~\ref{fig:f2}-(left), the data for $\pi^0$ and non-identified charged particles ($h^{\pm}$) are presented as the Nuclear Modification Factor, $R_{AA}(p_T)$, the ratio of the yield of $\pi^0$ (or $h^{\pm}$) per central Au+Au collision (upper 10\%-ile of observed multiplicity)  to the point-like-scaled p-p cross section:
   \begin{equation}
  R_{AA}(p_T)={{d^2N^{\pi}_{AA}/dp_T dy N_{AA}}\over {\langle T_{AA}\rangle d^2\sigma^{\pi}_{pp}/dp_T dy}} \quad , 
  \label{eq:RAA}
  \end{equation}
where $\mean{T_{AA}}$ is the overlap integral of the nuclear thickness functions. 
The $\pi^0$ data at nucleon-nucleon c.m. energy $\sqrt{s_{NN}}=200$~GeV are consistent with a constant $R_{AA}\sim 0.2$ over the range $4\leq p_T\leq 20$~GeV/c, while the suppression of non-identified charged hadrons and $\pi^0$ are different for $2\leq p_T \leq 6$~GeV/c and come together for $p_T > 6$~GeV/c. 
 
A new PHENIX result~\cite{ppg084} (Fig.~\ref{fig:f2}-(right)) nicely illustrates that parton suppression begins somewhere between $\sqrt{s_{NN}}$=22.4 and 62.4~GeV for Cu+Cu central collisions. This confirms that $\pi^0$ (jet) suppression is unique at RHIC energies and occurs at both $\sqrt{s_{NN}}=200$ and 62.4~GeV. The suppression is attributed to energy-loss of the outgoing hard-scattered color-charged partons due to interactions in the presumably deconfined and thus color-charged medium produced in Au+Au (and Cu+Cu) collisions at RHIC~\cite{BSZARNPS}. 
\subsubsection{Precise and accurate reference spectra are crucial}
\begin{figure}[!ht]
\begin{center}
\includegraphics[width=1.0\linewidth]{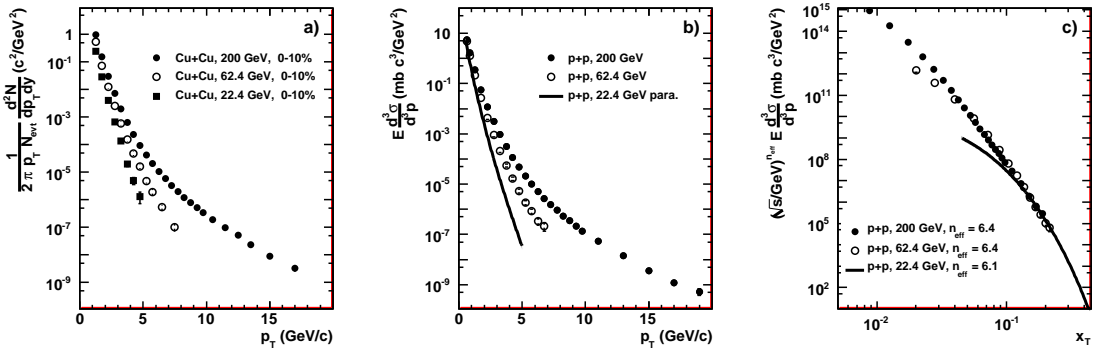} 
\end{center}\vspace*{-1.5pc}
\caption[]{Plots for $\sqrt{s_{NN}}$=22.4, 62.4 and 200~GeV of: a) measured invariant $\pi^0$ yields in central Cu+Cu collisions; b) measured invariant $\pi^0$ cross sections in p-p collisions at 62.4 and 200~GeV and fit at 22.4~GeV~\cite{Arleo-DdE}; c) the p-p data and fit from (b) plotted in the form $\sqrt{s}{\rm (GeV)}^{n_{\rm eff}} \times Ed^3\sigma/dp^3$ to exhibit $x_T$ scaling with $n_{\rm eff}$~6.1--6.4, consistent with Fig.~\ref{fig:otherxt}-(right)-b). }
\label{fig:xT22-200}
\end{figure}

    It is important to note that PHENIX did not measure the reference p-p spectrum at $\sqrt{s}=22.4$~GeV but used a QCD-based fit~\cite{Arleo-DdE} to the world's data on charged and neutral pions which was checked against PHENIX p-p measurements at 62.4 and 200~GeV using $x_T$ scaling (Fig.~\ref{fig:xT22-200})~\cite{ppg084}. 
A key issue in this fit is that the data at $\sqrt{s}=22.4$~GeV were consistent with each other and with pQCD~\cite{Arleo-DdE} except for one outlier which was excluded based on the experience from a previous global fit~\cite{DdE} to the world's data at $\sqrt{s}=62.4$~GeV where there are large disagreements (recall Fig.~\ref{fig:otherxt}-(left)-(top)). 
	The PHENIX measurement of the p-p reference spectrum at 62.4~GeV~\cite{ppg087} agreed with the measurements shown in Figs.~\ref{fig:ccorpt} and \ref{fig:otherxt}-(left)  to within the systematic error of the absolute $p_T$ scales, but disagreed significantly with the global fit at 62.4~GeV~\cite{DdE} which did not attempt to eliminate outliers and which had no basis for adjusting the absolute $p_T$ scales 
  \begin{figure}[!h]
\begin{center}
\begin{tabular}{cc}
\hspace*{-0.015\linewidth}\includegraphics[width=0.48\linewidth]{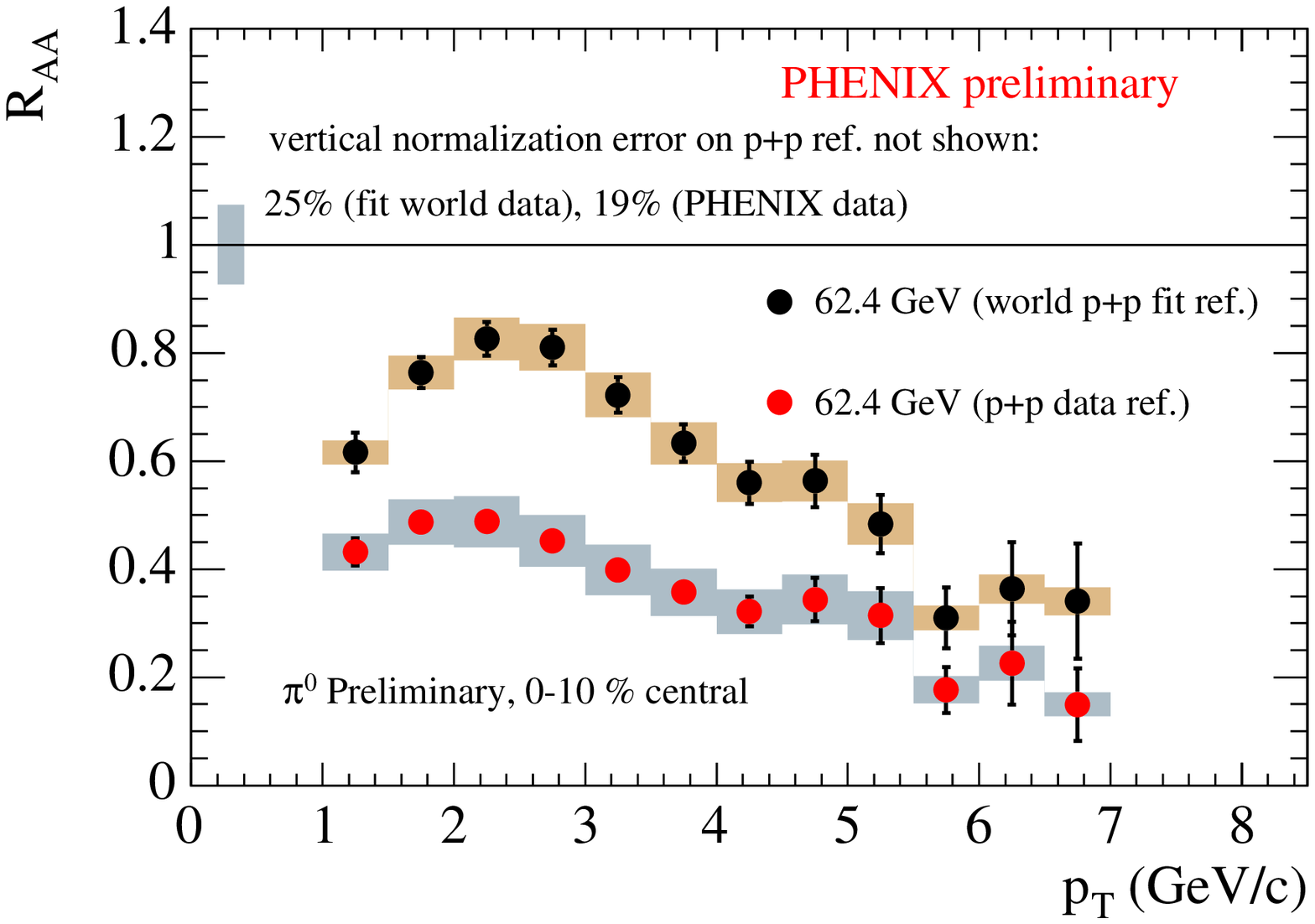} & 
\includegraphics[width=0.48\linewidth]{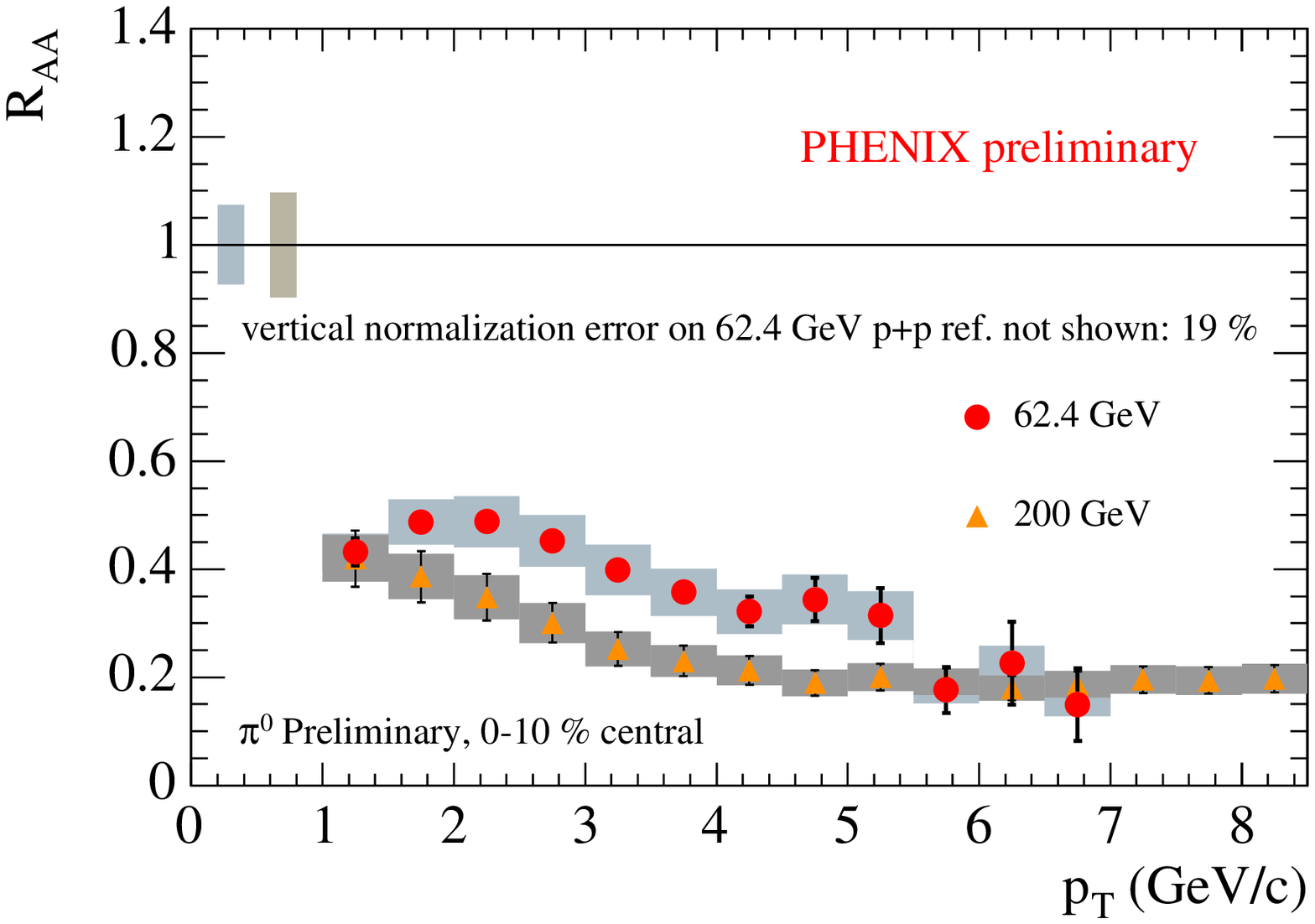}  
\end{tabular}
\end{center}\vspace*{-1.5pc}
\caption[]{(left) Comparison of $R_{AA}(p_T)$ for $\pi^0$ in $\sqrt{s_{NN}}=62.4$~GeV central Au+Au collisions using the fit~\cite{DdE} to the previous world 62.4~GeV p-p data or the measured PHENIX reference 62.4~GeV p-p data~\cite{ppg087}. (right) Final $R_{AA}(p_T)$ in central Au+Au collisions at $\sqrt{s_{NN}}=62.4$ and 200~GeV~\cite{Gabor08}.}
\label{fig:RAA2ways}
\end{figure}		
of the various measurements.
	In Fig.~\ref{fig:RAA2ways}-(left), the $R_{AA}(p_T)$ for Au+Au central collisions 
at $\sqrt{s_{NN}}=62.4$~GeV computed with the global p-p fit~\cite{DdE} and the measured reference spectrum~\cite{ppg087} are shown,  while the final $R_{AA} (p_T)$ for Au+Au central collisions at 62.4 is compared to $R_{AA}(p_T)$ at 200~GeV in Fig.~\ref{fig:RAA2ways}-(right)~\cite{Gabor08}. 
    If it weren't already obvious, this should be a lesson to the LHC physicists (and management) of the importance of making reference measurements in the same detector for  p-p collisions at the identical $\sqrt{s}$ as the $\sqrt{s_{NN}}$ of the A+A collisions. 
\subsection{$J/\Psi$-suppression---still golden?}
\begin{figure}[!ht]
\begin{center}
\begin{tabular}{cc}
\hspace*{-0.015\linewidth}\includegraphics[width=0.48\linewidth,height=0.6\linewidth]{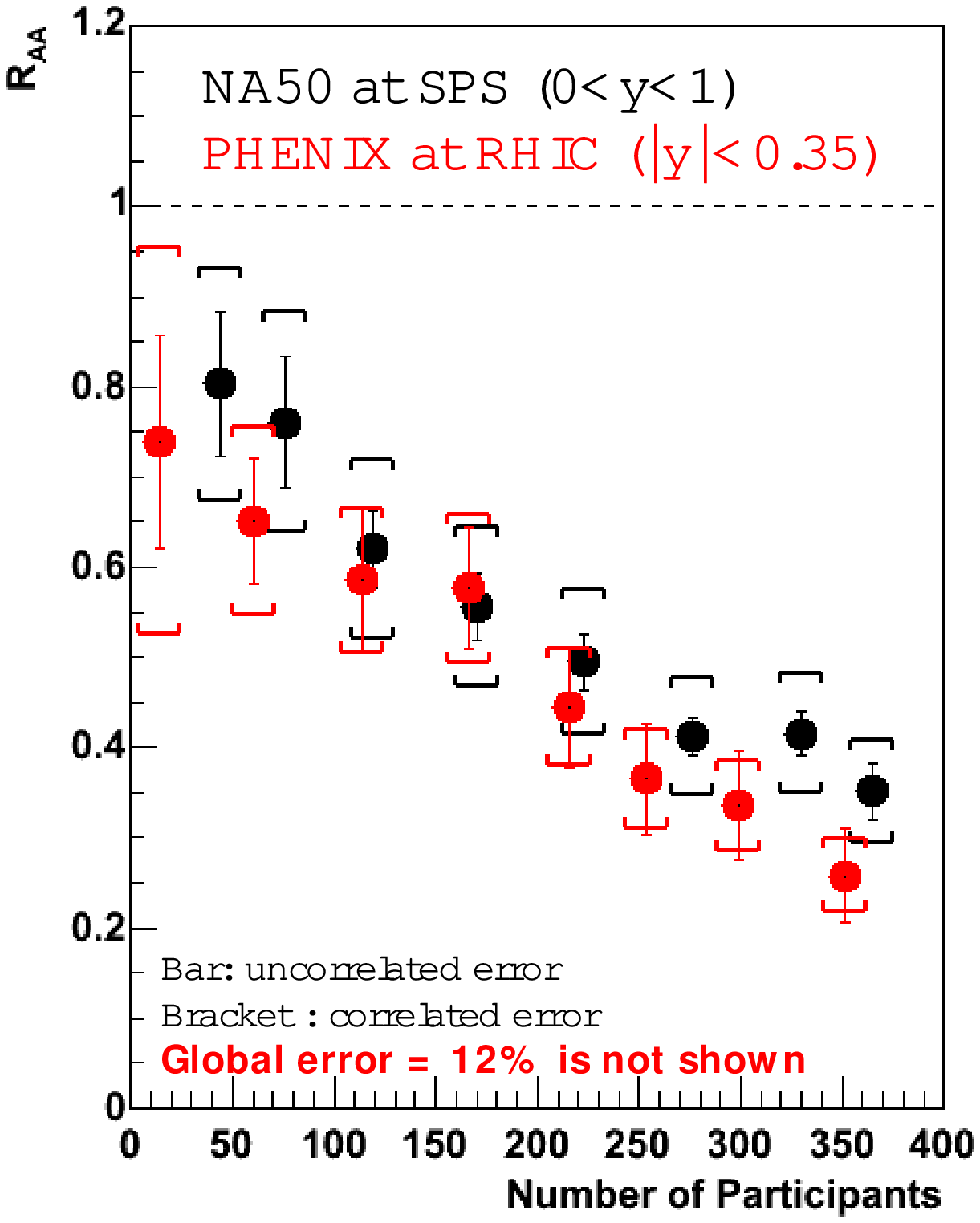} & 
\includegraphics[width=0.48\linewidth,height=0.6\linewidth]{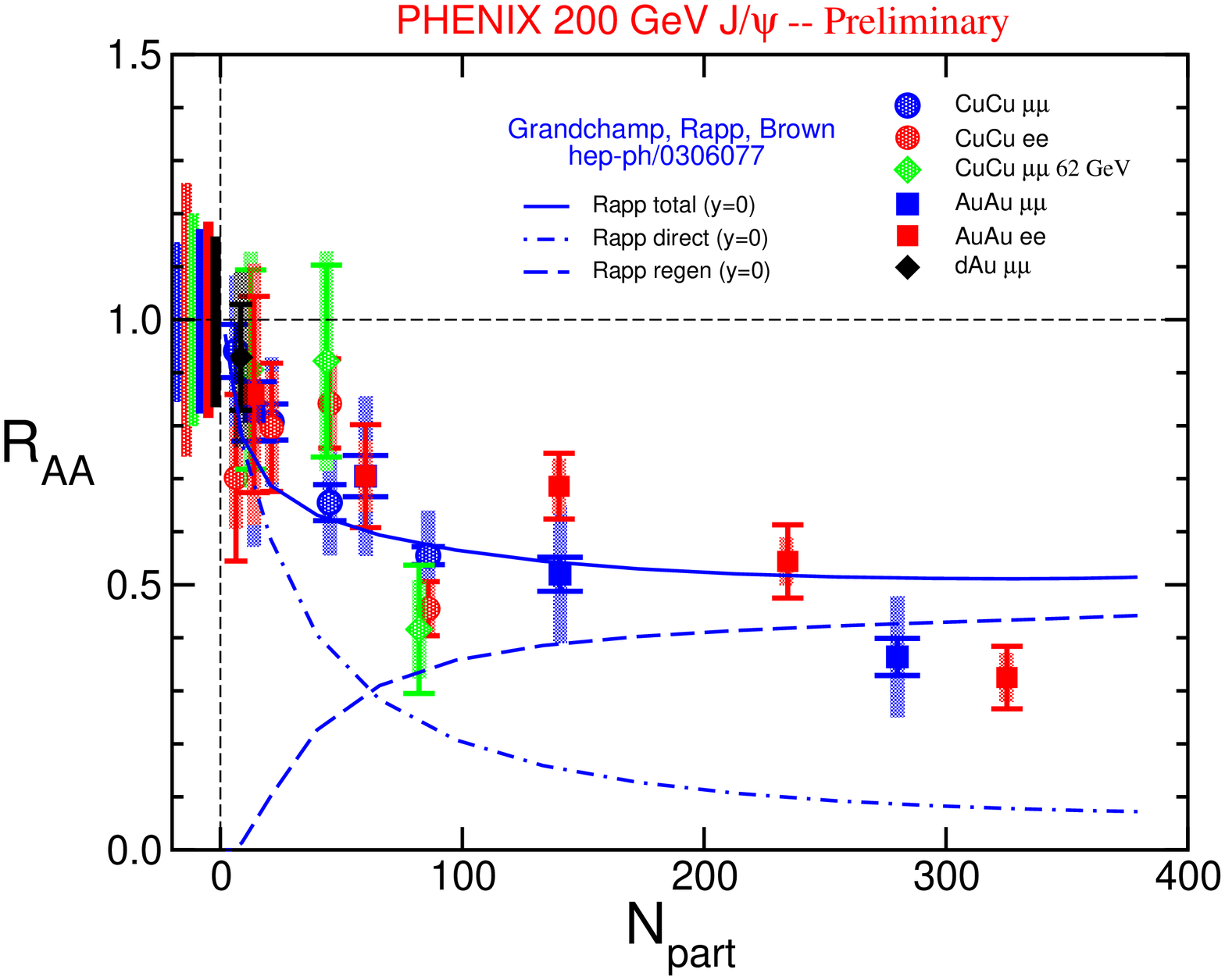}  
\end{tabular}
\end{center}
\caption[]{(left) $R_{AA}^{J/\Psi}$ vs centrality ($N_{\rm part}$) at RHIC and SpS energies~\cite{PRL98-1}. (right) Predictions for $R_{AA}^{J/\Psi}$ in a model with regeneration~\cite{GRB}.}
\label{fig:f3}
\end{figure}

The dramatic difference in suppression of hard-scattering at RHIC compared to  SPS fixed target c.m. energy ($\sqrt{s_{NN}}=17$~GeV) stands in stark contrast to $J/\Psi$ suppression, originally thought to be the gold-plated signature for deconfinement and the Quark Gluon Plasma (QGP)~\cite{MatsuiSatz}. $R_{AA}$ for $J/\Psi$ suppression is the same, if not identical, at SPS and RHIC (see Fig.~\ref{fig:f3}-(left)), thus casting a serious doubt on the value of $J/\Psi$ suppression as a probe of deconfinement. The medium at RHIC makes $\pi^0$'s  nearly vanish but leaves the $J/\Psi$ unchanged compared to lower $\sqrt{s_{NN}}$. One possible explanation is that $c$ and $\bar{c}$ quarks in the QGP recombine to regenerate $J/\Psi$ (see Fig.~\ref{fig:f3}-(right)), miraculously making the observed $R_{AA}$ equal at SpS and RHIC c.m. energies. The good news is that such models predict $J/\Psi$ enhancement ($R_{AA}> 1$) at LHC energies, which would be spectacular, if observed.       

	This leaves us with the interesting question: Will Peter Higgs or Helmut Satz have to wait longer at LHC to find out whether they are right?
	
\section{The baryon anomaly and $x_T$ scaling} 
   Many RHI physicists tend to treat non-identified charged hadrons $h^{\pm}$ as if they were $\pi^{\pm}$. While this may be a reasonable assumption in p-p collisions, it is clear from Fig.~\ref{fig:f2}-(left) that the suppression of non-identified charged hadrons and $\pi^0$ is very different for $1 < p_T \leq 6$~GeV/c. 
   
\begin{figure}[tbhp]
   \begin{center}
\includegraphics[width=0.8\linewidth]{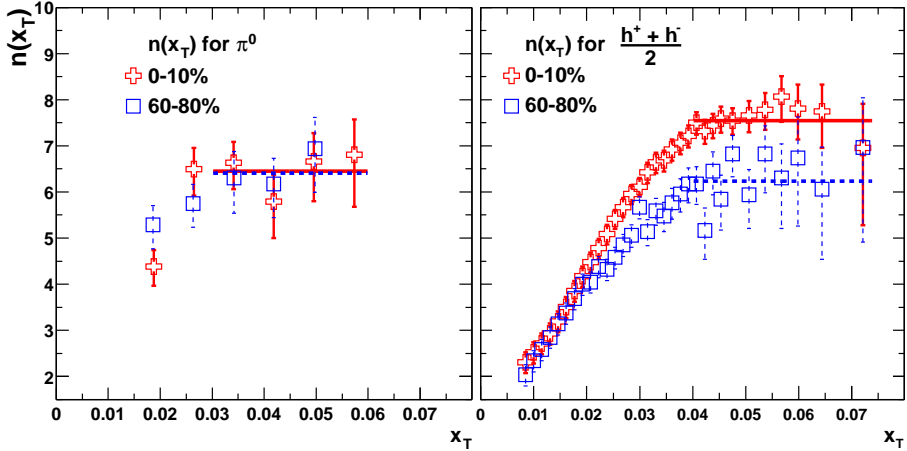}
\end{center}
	\vspace*{-0.24in}
\caption[]{Power-law exponent $n_{\rm eff}(x_T)$ for $\pi^0$ and $h^{\pm}$ spectra in central and peripheral Au+Au collisions at $\sqrt{s_{NN}} = 130$ and 200~GeV~\cite{PXxTAuAu}. } 
\label{fig:nxTAA}
\end{figure}

	If the production of high-$p_T$ particles in Au+Au collisions is the
result of hard scattering according to pQCD, then $x_T$ scaling should
work just as well in Au+Au collisions as in p-p collisions and should
yield the same value of the exponent $n_{\rm eff}(x_T,\sqrt{s})$.  The only assumption
required is that the structure and fragmentation functions in Au+Au
collisions should scale, in which case Eq. \ref{eq:nxt} still
applies, albeit with a $G(x_T)$ appropriate for Au+Au. In
Fig.~\ref{fig:nxTAA}, $n_{\rm eff}(x_T,\sqrt{s_{NN}})$ in Au+Au is shown for $\pi^0$ and $h^{\pm}$ in peripheral and central collisions, derived by
taking the ratio of $E d^3\sigma/dp^3$ at a given $x_T$ for
$\sqrt{s_{NN}} = 130$ and 200~GeV, in each case.~\cite{PXxTAuAu}  

The $\pi^0$'s exhibit
$x_T$ scaling, with the same value of $n_{\rm eff} = 6.3$ as in p-p collisions,
for both Au+Au peripheral and central collisions. The $x_T$ scaling establishes
that high-$p_T$ $\pi^0$ production in peripheral and central Au+Au
collisions follows pQCD as in p-p collisions, with parton distributions and 
fragmentation functions that scale with $x_T$, at least within
the experimental sensitivity of the data. The fact that the fragmentation functions scale for $\pi^0$ in Au+Au central collisions indicates that the effective energy loss must scale, i.e. $\Delta E(p_T)/p_T$ is a constant, which is  consistent with the constant value of $R_{AA}(p_T)$ for $p_T>4$~GeV/c (Fig.~\ref{fig:f2}-(left)), given that the $\pi^0$ $p_T$ spectrum is a pure power-law (Fig.~\ref{fig:otherxt}-(right)-a)). 

\begin{figure}[!thb]
\begin{center}
\begin{tabular}{cc}
\includegraphics[width=0.52\linewidth]{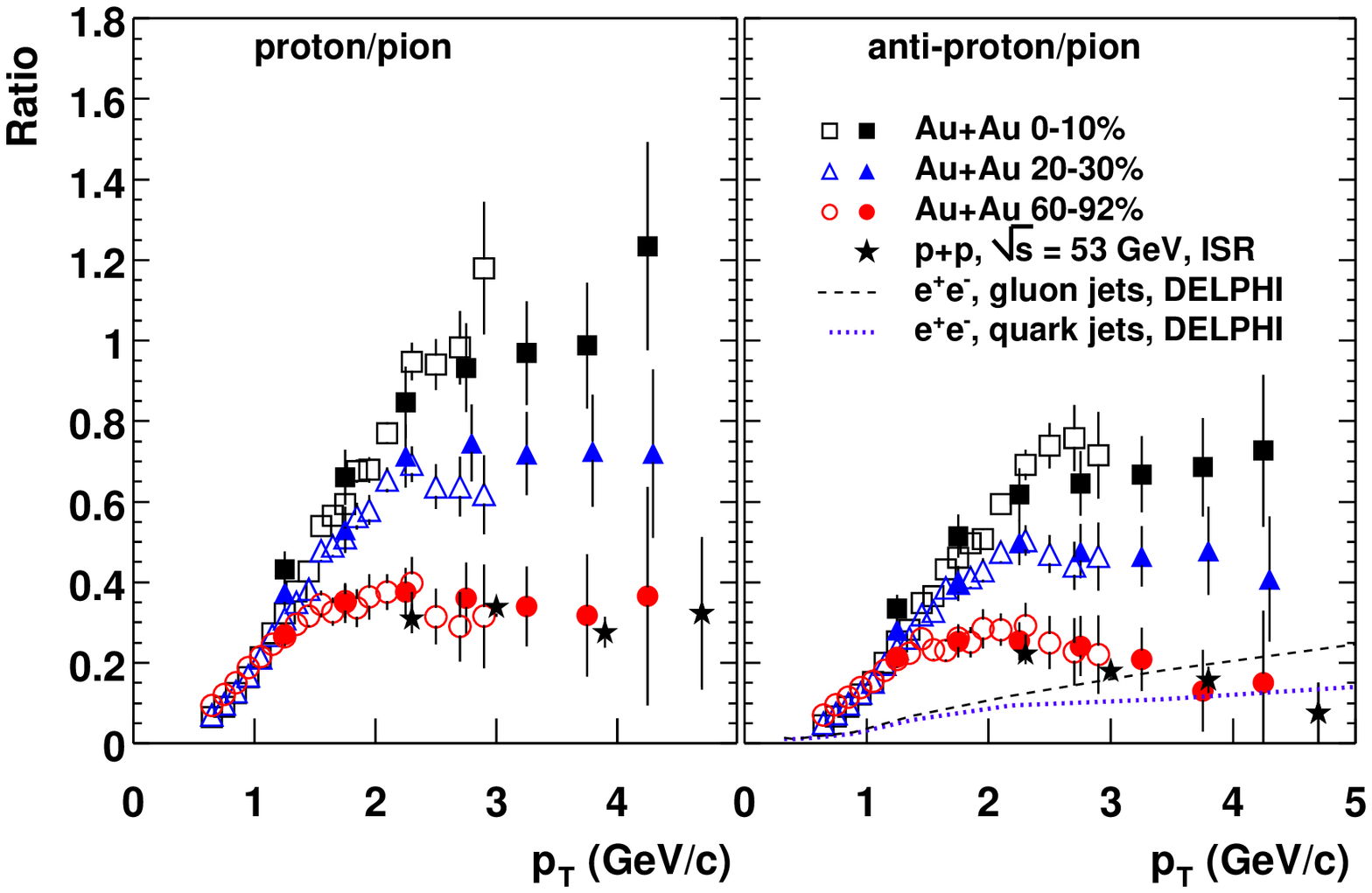}&
\hspace*{-0.04\linewidth}\includegraphics[width=0.52\linewidth]{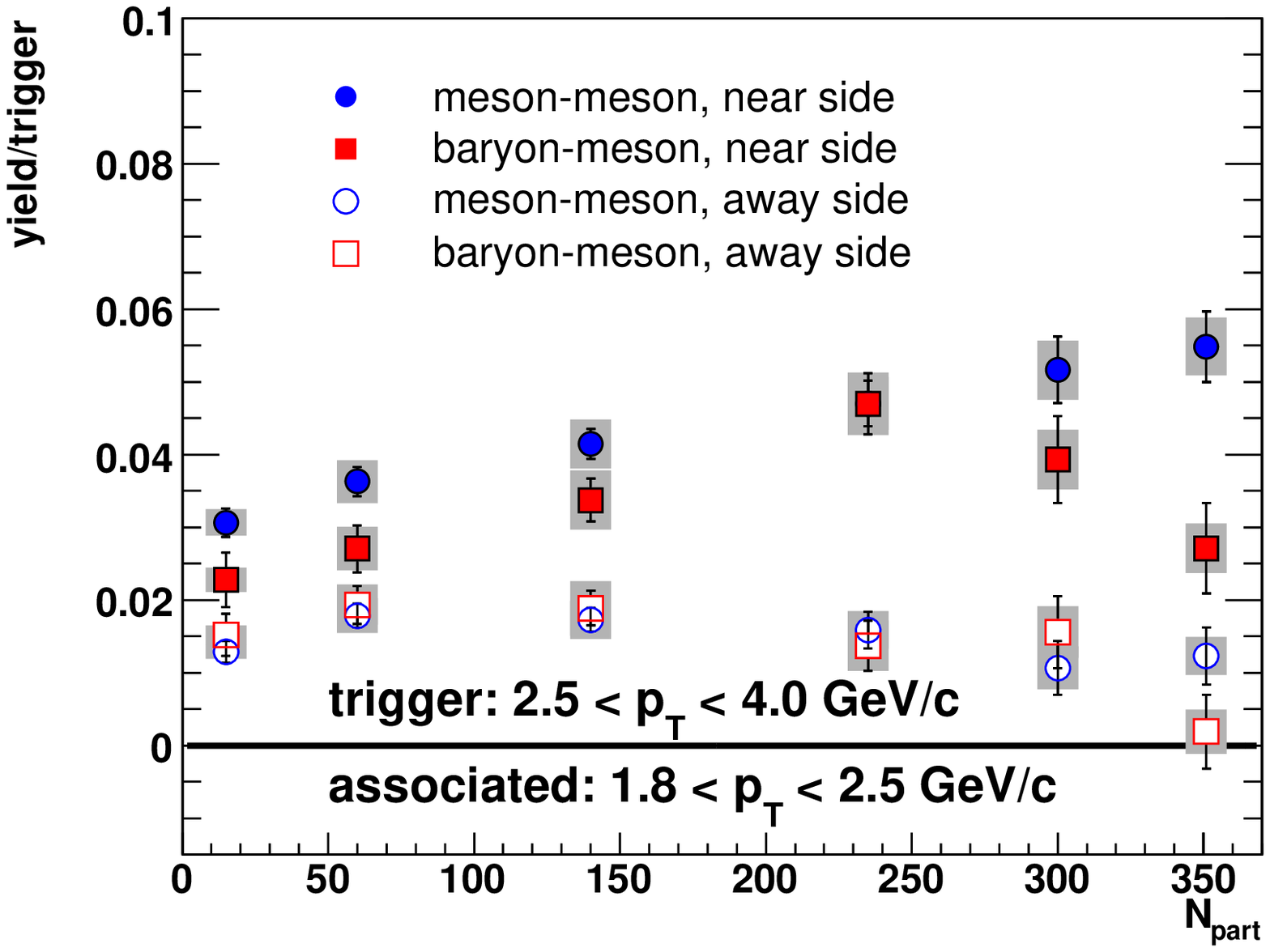}
\end{tabular}
\end{center}\vspace*{-0.25in}
\caption[]{(left) $p/\pi^+$ and $\bar{p}/\pi^-$ as a function of $p_T$ and centrality from Au+Au collisions at $\sqrt{s_{NN}}=200$~GeV~\cite{ppg015} compared to other data indicated; (right) Conditional yields, per trigger meson (circles), baryon (squares) with $2.5< p_T < 4$~GeV/c, of associated mesons with $1.7 < p_T < 2.5$~GeV/c integrated within $\Delta\phi=\pm 0.94$ radian of the trigger (near side-full) or opposite azimuthal angle (open), for Au+Au collisions at $\sqrt{s_{NN}}=200$~GeV~\cite{ppg072}. }
\label{fig:banomaly}
\end{figure}

    The deviation of $h^{\pm}$ in Fig.~\ref{fig:nxTAA} from $x_T$ scaling in central Au+Au collisions is indicative of and consistent with the strong non-scaling modification of particle composition of identified hadrons observed in Au+Au collisions compared to that of p-p collisions in the range $2.0\leq p_T\leq 4.5$~GeV/c, where particle production is the result of jet-fragmentation. This is called the Baryon Anomaly. As shown in Fig.~\ref{fig:banomaly}-(left) the $p/\pi^{+}$ and $\bar{p}/\pi^{-}$ ratios as a function of $p_T$ increase dramatically to values $\sim$1 as a function of centrality in Au+Au collisions at RHIC~\cite{ppg015}. This is nearly an order of magnitude larger than had ever been seen previously in either fragmentation of jets in $e^+ e^-$ collisions or in the average particle composition of the bulk matter in Au+Au central collisions~\cite{ppg026}.  
    
   This `baryon anomaly' was beautifully explained as due to the coalescence of an exponential (thermal) distribution of constituent quarks (a.k.a. the QGP)~\cite{GFH03}. Unfortunately, measurements of correlations of $h^{\pm}$ in the range $1.7\leq p_{T_a}\leq 2.5$~GeV/c associated to identified meson or baryon triggers  with $2.5\leq p_{T_t}\leq 4.0$~GeV/c showed the same near side and away side peaks and yields (Fig.~\ref{fig:banomaly}-(right)) characteristic of di-jet production from hard-scattering~\cite{PXPRC71,ppg072}, rather than from soft coalescence, apparently ruling out this beautiful model. 
   
   There are still plenty of other models of the baryon anomaly, but none of them are clearly definitive. For instance, Stan Brodsky presented at this meeting~\cite{StanAnne} a higher twist model of the baryon anomaly as the result of the reaction $q+q\rightarrow p+\bar{q}$. This predicts an isolated proton with no same-side jet, but with an opposite jet, a clear and crucial test. Another test (from the CIM~\cite{CIM}) is that $n_{\rm eff}\rightarrow 8$ for these protons. This effect will be emphasized in central collisions because the higher twist subprocesses have `small size' and are `color transparent' so they propagate through the nuclear medium without absorption. This is consistent with the reduced near-side correlation to baryon triggers compared to meson triggers shown for the most central collisions in Fig.~\ref{fig:banomaly}-(right)~\cite{ppg072}; but definitive detection of isolated $p$ or $\bar{p}$ in central A+A collisions or precision measurements of $x_T$ scaling for $p$ and $\bar{p}$ as a function of centrality remain very interesting projects for the future. 
   
   Guy Paic recently~\cite{CuautlePaic08} claimed to explain the baryon anomaly by simple radial flow, which occurs late in the expansion, even in the hadronic phase. The radial flow velocity Lorentz-boosts the heaver protons to larger $p_T$ than the lighter pions. Also, protons have a shorter formation time than pions so they may participate in the radial flow even if they result from parton fragmentation. If Guy is correct, then it is time to re-examine the subject of the elliptic flow ($v_2$) and $p_T$ spectra of identified hadrons which was very popular several years ago. The last time I looked, in the PHENIX White Paper~\cite{PXWP}, the hydro models with radial flow could either explain the $p_T$ spectra of $\bar{p}$ and $\pi$ or the $v_2$ but not both. If the steadily improving models can now explain both $v_2$ and the $p_T$ spectra, this would spell the end of the `baryon anomaly'. However, the anomalously large $\bar{p}/\pi$ ratio would remain a signature property of the inclusive identified particle $p_T$ spectra in central Au+Au collisions. 
         
   In this vein, I was temporarily thrown for a loop by a result from a STAR presentation at Quark Matter 2008, as reported by Marco van Leeuwen at Hard Probes 2008~\cite{MarcoHP08} (and Christine Nattrass~\cite{Nattthis} at this meeting), which measured the particle composition in the near-side jet and `ridge' and seemed to indicate that the $p/\pi$ (or baryon/meson) ratio in the near-side jet was not anomalous. I first thought that this disagreed with everything that I said previously in this section about the `baryon anomaly'.  To quote Marco~\cite{MarcoHP08}, ``the $p/\pi$ ratio in the ridge is similar to the inclusive $p/\pi$ ratio in Au+ Au events, which is much larger than in p+p events. The $p/\pi$ ratio in the jet-like peak is similar to the inclusive ratio in p + p events.'' However, at a Ridge workshop at BNL~\cite{Ridgewks}, I found out that what STAR really meant to say was that ``The $p/\pi$ ratio of the conditional yield for near side-correlations associated to an $h^{\pm}$ trigger with $p_{T_t}>4$~GeV/c is similar in the jet-like peak to the inclusive ratio in p + p events; while the $p/\pi$ ratio in the ridge is similar to the inclusive $p/\pi$ ratio in Au+ Au events, which is much larger than in p+p events.''---i.e. STAR was talking about associated yields to the trigger $h^{\pm}$ and did not include the triggering particle in the yield. Hence there is no disagreement. In fact, the STAR result is actually in agreement with a recent PHENIX measurement~\cite{ppg034} of the ratio of the associated baryon and meson conditional near-side yields from an $h^{\pm}$ trigger with $2.5< p_{T_t}< 4$~GeV/c in Au+Au collisions at $\sqrt{s_{NN}}=200$~GeV.   
   
   To reiterate, the anomalously large $\bar{p}/\pi$ ratio remains a signature property of the inclusive identified particle $p_T$ spectra in central Au+Au collisions. In fact, Christine Nattrass'~\cite{Nattthis} thorough demonstration at this meeting of the properties and particle composition of `the ridge' convinced me that that the ridge is nothing other than a random coincidence of any trigger particle and the background or bulk of inclusive particles, which, as Stan Brodsky commented, is generally biased towards the trigger direction due to the $k_T$ effect. Naturally, several details such as the azimuthal width of the ridge still remain to be explained. In my opinion, a key test of this idea is that a ridge of same side correlations with large $\Delta\eta$ should exist for direct-$\gamma$ triggers, or, in the PHENIX acceptance, the same-side correlation to a direct-$\gamma$ should exist at the same rate and azimuthal width as the ridge we observed in $\pi^0$ or inclusive $\gamma$ same-side correlations.~\cite{ppg083}  
   
\section{Direct photons at RHIC---Thermal photons?}
\subsection{Internal Conversions---the first measurement anywhere of direct photons at low $p_T$}
\begin{figure}[!h]
\begin{center}
\begin{tabular}{cc}
\includegraphics[width=0.55\linewidth]{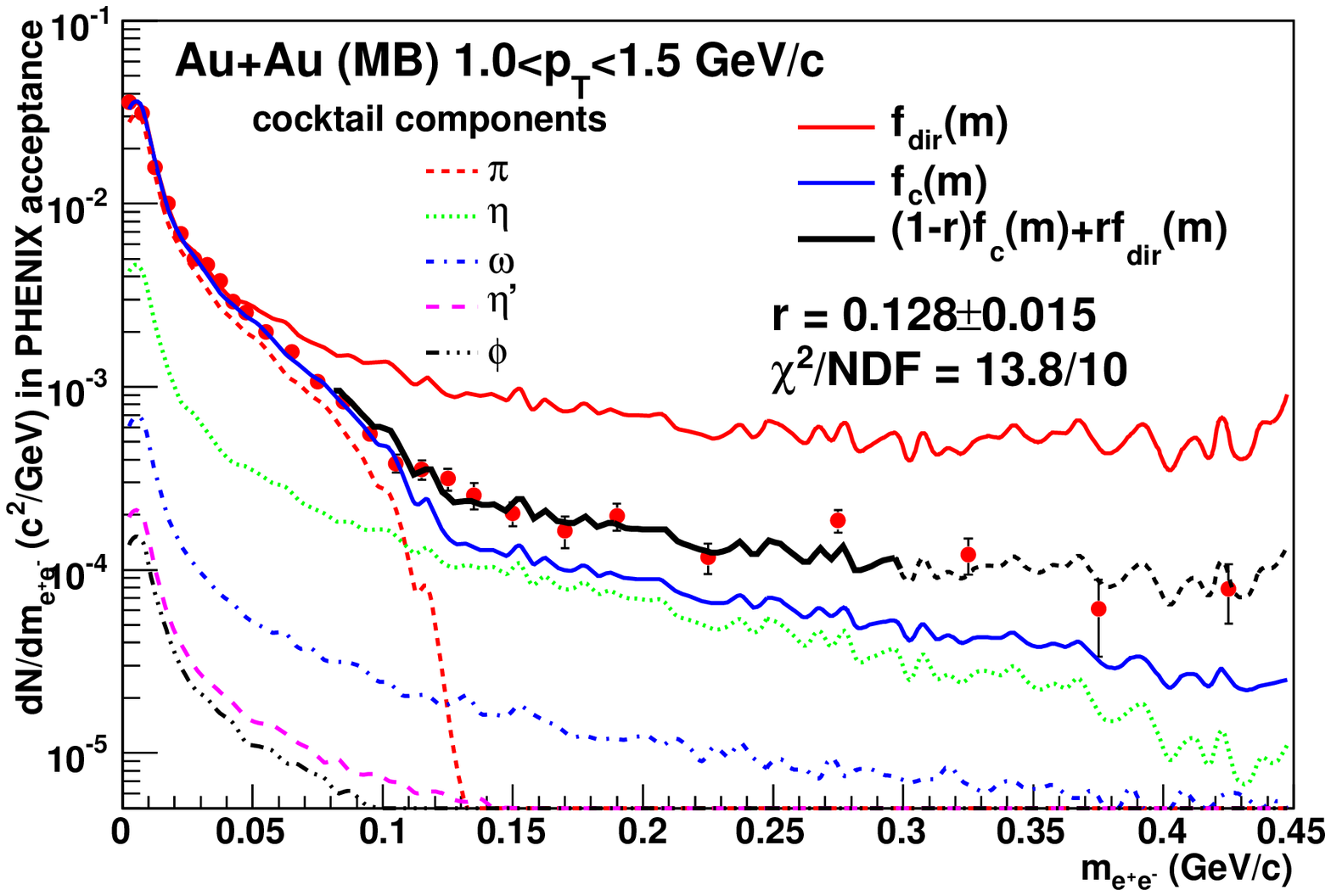}& 
\hspace*{-0.02\linewidth}\includegraphics[width=0.45\linewidth]{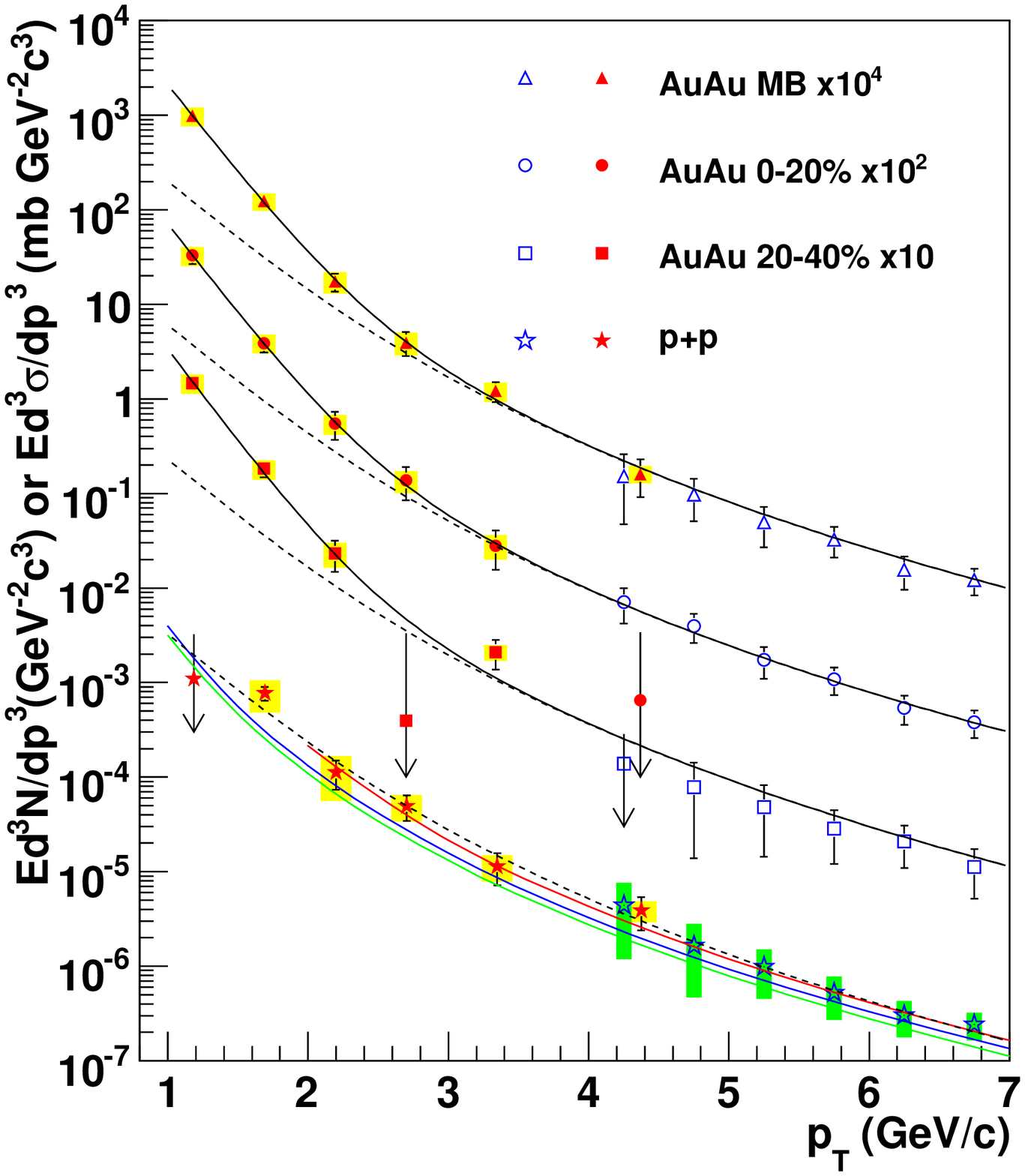} 
\end{tabular}
\end{center}
\caption[]{(left) Invariant mass ($m_{e+ e^-}$) distribution of $e^+ e^-$ pairs from Au+Au minimum bias events for $1.0< p_T<1.5$~GeV/c~\cite{ppg086}. Dashed lines are Eq.~\ref{eq:KW} for the mesons indicated. Blue solid line is $f_c(m)$, the total di-electron yield from the sum of contributions or `cocktail' of meson Dalitz decays; Red solid line is $f_{dir}(m)$ the internal conversion $m_{e^+ e^-}$ spectrum from a direct-photon ($M>> m_{e^+ e^-}$). Black solid line is a fit of the data to the sum of cocktail plus direct contributions in the range $80< m_{e+ e^-} < 300$ MeV/c$^2$. (right) Invariant cross section (p-p) or invariant yield (Au+Au) of direct photons as a function of $p_T$~\cite{ppg086}. Filled points are from virtual photons, open points from real photons.  
\label{fig:ppg086Figs} }
\end{figure}

   Internal conversion of a photon from $\pi^0$ and $\eta$ decay is well-known and is called Dalitz decay~\cite{egNPS}. Perhaps less well known in the RHI community is the fact that for any reaction (e.g. $q+g\rightarrow \gamma +q$) in which a real photon can be emitted, a virtual photon (e.g. $e^+ e^-$ pair of mass $m_{ee}\geq 2m_e$) can also be emitted. This is called internal-conversion and is generally given by the Kroll-Wada formula~\cite{KW,ppg086}:
   \begin{eqnarray}
   {1\over N_{\gamma}} {{dN_{ee}}\over {dm_{ee}}}&=& \frac{2\alpha}{3\pi}\frac{1}{m_{ee}} (1-\frac{m^2_{ee}}{M^2})^3 \quad \times \cr & &|F(m_{ee}^2)|^2 \sqrt{1-\frac{4m_e^2}{m_{ee}^2}}\, (1+\frac{2m_e^2}{m^2_{ee}})\quad ,
   \label{eq:KW}
   \end{eqnarray}
   where $M$ is the mass of the decaying meson or the effective mass of the emitting system. The dominant terms are on the first line of Eq.~\ref{eq:KW}:  the characteristic $1/m_{ee}$ dependence; and the cutoff of the spectrum for $m_{ee}\geq M$ (Fig.~\ref{fig:ppg086Figs}-(left))~\cite{ppg086}. Since the main background for direct-single-$\gamma$ production is a photon from $\pi^0\rightarrow \gamma +\gamma$, selecting $m_{ee} \gsim 100$ MeV effectively reduces the background by an order of magnitude by eliminating the background from $\pi^0$ Dalitz decay, $\pi^0\rightarrow \gamma + e^+ + e^- $, at the expense of a factor $\sim 1000$ in rate. This allows the direct photon measurements to be extended (for the first time in both p-p and Au+Au collisions) below the value of $p_T\sim 4$~GeV/c, possible with real photons, down to $p_T=1$~GeV/c (Fig.~\ref{fig:ppg086Figs}-(right))~\cite{ppg086}, which is a real achievement. 
The solid lines on the p-p data are QCD calculations which work down to $p_T=2$~GeV/c. The dashed line is a fit of the p-p data to the modified power law $B (1+p_T^2/b)^{-n}$, used in the related Drell-Yan~\cite{Ito81} reaction, which flattens as $p_T\rightarrow 0$. 

	The relatively flat, non-exponential, spectra for the direct-$\gamma$ and Drell-Yan reactions as $p_T\rightarrow 0$ is due to the fact that there is no soft-physics production process for them, only production via the partonic subprocesses, $g+q\rightarrow \gamma+q$ and $\bar{q}+q\rightarrow e^+ + e^-$, respectively.  This is quite distinct from the case for hadron production, e.g. $\pi^0$, where the spectra are exponential  as $p_T\rightarrow 0$ in p-p collisions (Fig.~\ref{fig:PXpi0pp}) due to soft-production processes, as well as in Au+Au collisions. 
 Thus, for direct-$\gamma$ in  Au+Au collisions, the exponential spectrum of excess photons above the $\mean{T_{AA}}$ extrapolated p-p fit is unique and therefore suggestive of a thermal source.   
\subsection{Low $p_T$ vs high $p_T$ direct-$\gamma$---Learn a lot from a busy plot}

\begin{figure}[!h]
\begin{center}
\includegraphics[width=0.90\linewidth]{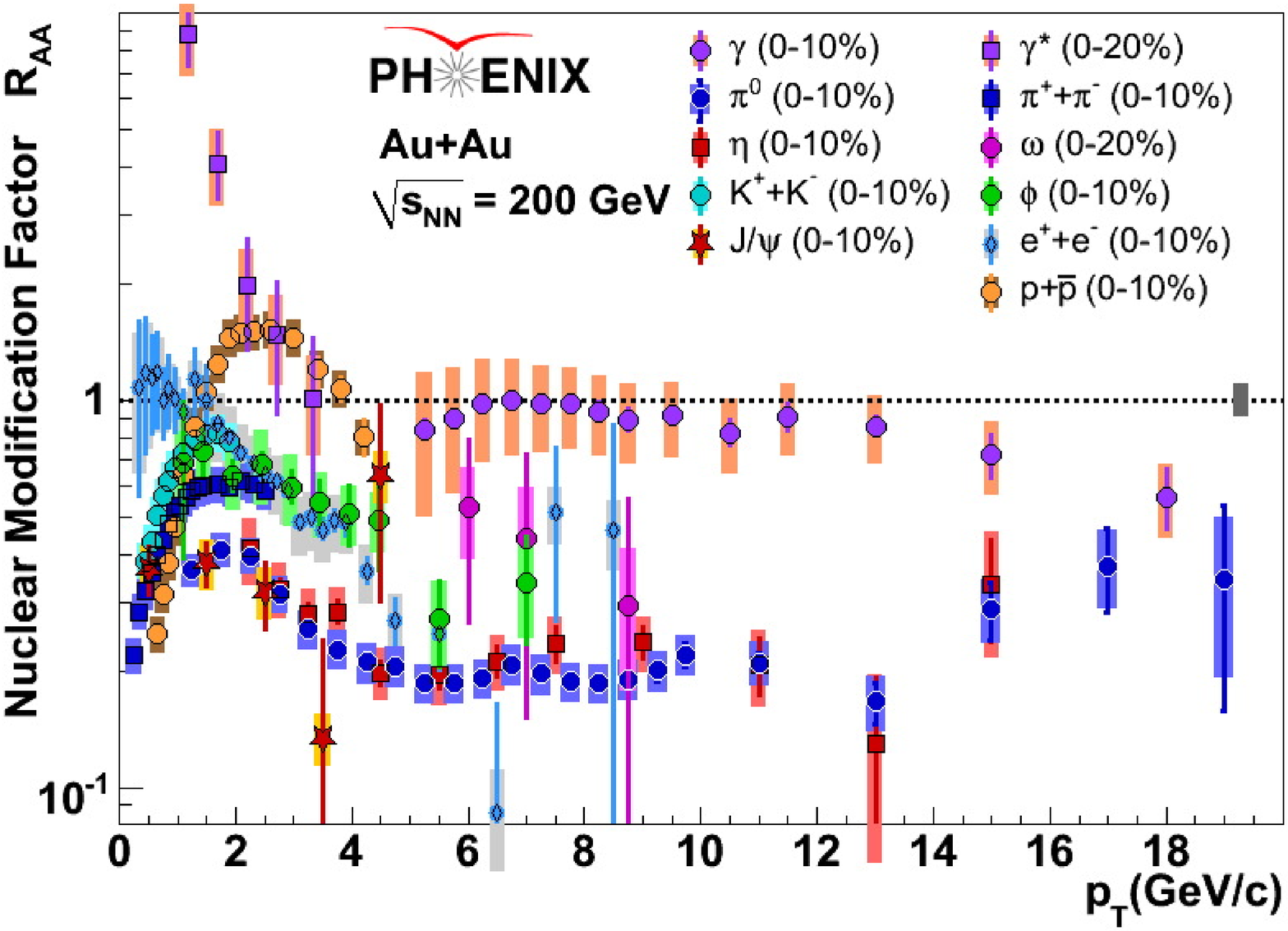} 
\end{center}
\caption[]{Nuclear Modification Factor, $R_{AA}(p_T)$ for all identified particles so far measured by PHENIX in central Au+Au collisions at $\sqrt{s_{NN}}=200$~GeV.~\cite{ThanksAM}   
\label{fig:Tshirt} }
\end{figure}

     The unique behavior of direct-$\gamma$ at low $p_T$ in Au+Au relative to p+p compared to any other particle is more dramatically illustrated by examining  the $R_{AA}$ of all particles measured by PHENIX in central Au+Au collisions at $\sqrt{s_{NN}}=200$~GeV (Fig.~\ref{fig:Tshirt})~\cite{ThanksAM}. For the entire region $p_T\leq 20$~GeV/c so far measured at RHIC, apart from the $p+\bar{p}$ which are enhanced in the region $2\leq p_T \lsim 4$~GeV/c ('the baryon anomaly'), the production of {\em no other particle} is enhanced over point-like scaling. 
The behavior of $R_{AA}$ of the low $p_T\leq 2$~GeV/c direct-$\gamma$ is totally and dramatically different from all the other particles, exhibiting an order of magnitude exponential enhancement as $p_T\rightarrow 0$. This exponential enhancement is certainly suggestive of a new production mechanism in central Au+Au collisions different from the conventional soft and hard particle production processes in p-p collisions and its unique behavior is attributed to thermal photon production by many authors.~\cite{DdE-DP}

\subsubsection{Direct photons and mesons up to $p_T=20$~GeV/c}
   Other instructive observations can be gleaned from Fig.~\ref{fig:Tshirt}.  The $\pi^0$ and $\eta$ continue to track each other to the highest $p_T$. At lower $p_T$, the $\phi$ meson tracks the $K^{\pm}$ very well, but with a different value of $R_{AA}(p_T)$ than the $\pi^0$, while at higher $p_T$,the $\phi$ and $\omega$ vector mesons appear to track each other. Interestingly, the $J/\Psi$ seems to track the $\pi^0$ for $0\leq p_T\leq 4$~GeV/c; and it will be interesting to see whether this trend continues at higher $p_T$. 
   
   The direct-$\gamma$'s also show something interesting at high $p_T$ which might possibly indicate trouble ahead at the LHC. With admittedly large systematic errors, which should not be ignored, the direct-$\gamma$ appear to become suppressed for $p_T> 14$~GeV/c with a trend towards equality with $R_{AA}^{\pi^0}$ for $p_T\sim 20$~GeV. Should $R_{AA}^{\gamma}$ become equal to $R_{AA}^{\pi^0}$, it would imply that the energy loss in the final state is no longer a significant effect for $p_T\gsim 20$~GeV/c and that the equal suppression of direct-$\gamma$ and $\pi^0$ is due to the initial state structure functions. If this were true, it could mean that going to much higher $p_T$ would not be useful for measurements of parton suppression. In this vein, the new EPS09 structure functions for quarks and gluons in nuclei were presented at this meeting~\cite{EPS09}, which represented the best estimate of shadowing derived by fitting all the  DIS data in $\mu(e)-A$ scattering as well as including in the fit, notably, the PHENIX $\pi^0$ data in d+Au and p-p as a function of centrality.  Clearly, improved measurements of both direct-$\gamma$ and $\pi^0$ in the range $10<p_T<20$~GeV/c are of the utmost importance for both the RHIC and LHC programs.      

\section{Precision measurements, key to the next step in understanding}
\begin{figure}[!h] 
\begin{center}
\begin{tabular}{cc}
\hspace*{-0.02\linewidth}\includegraphics[width=0.74\linewidth]{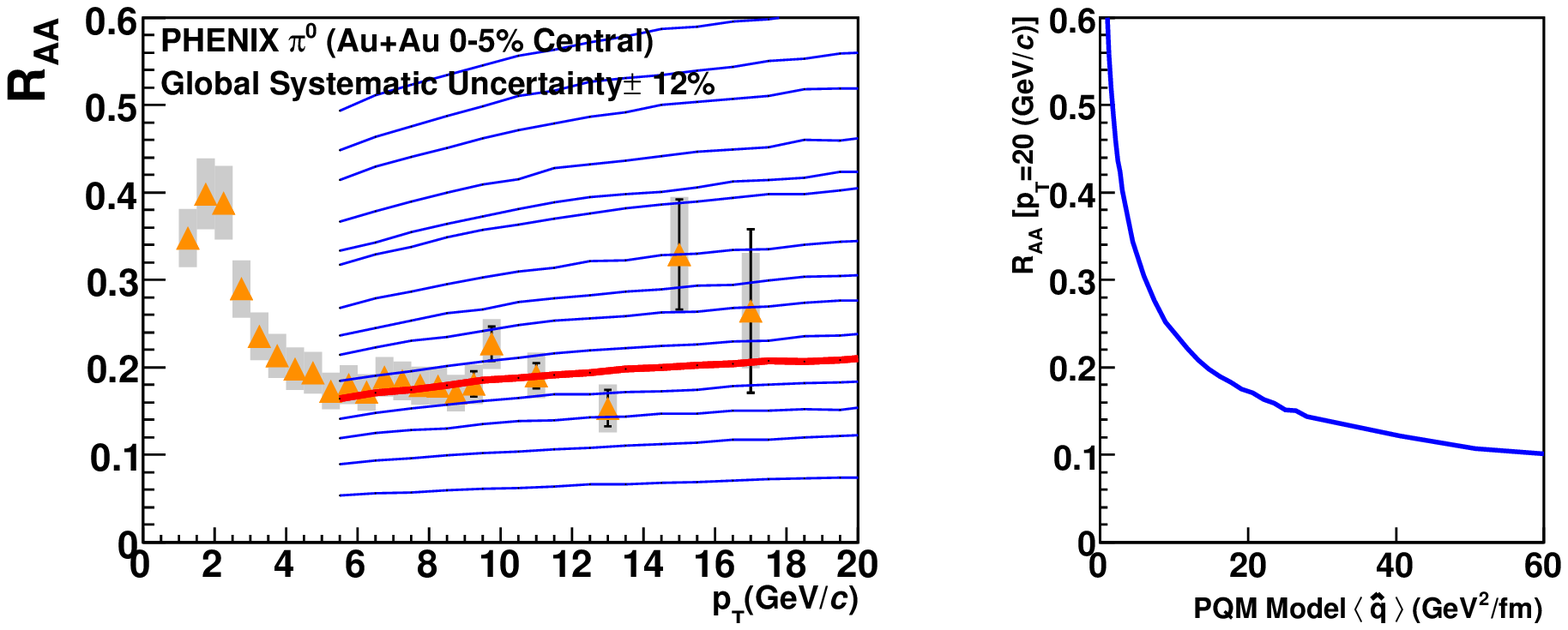} &
\hspace*{-0.03\linewidth}\includegraphics[width=0.29\linewidth]{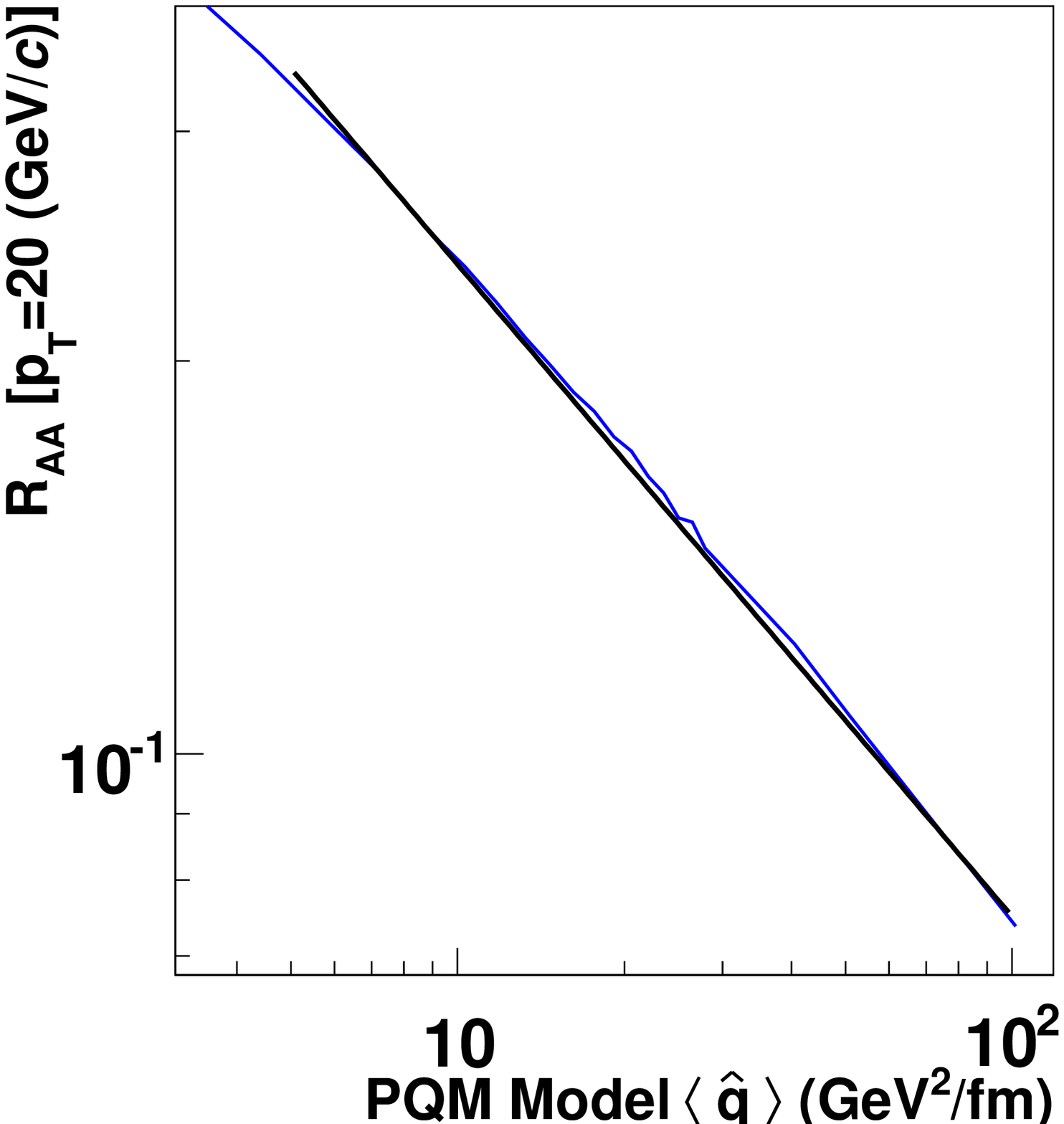}
\end{tabular}
\end{center}
\caption[]
{a) (left) PHENIX $\pi^0$ $R_{AA}(p_T)$ for Au+Au central (0-5\%) collisions at $\sqrt{s_{NN}}=200$~\cite{ppg079} compared to PQM model predictions~\cite{PQM}  as a function of $\mean{\hat{q}}$. The thick red line is the best fit. b) (center) Values of $R_{AA}$ at $p_T=20$~GeV/c as a function of $\mean{\hat{q}}$ in the PQM model~\cite{PQM} corresponding to the lines the left panel. c) (right) same as b) but on a log-log scale, with fit.       
\label{fig:pi0pqm} }
\end{figure}

   There are many different models of parton suppression with totally different assumptions which all give results in agreement with the PHENIX measurement $R_{AA}^{\pi^0}\approx 0.20$ for $4\leq p_T\leq 20$~GeV/c in Au+Au central collisions. In PHENIX, Jamie Nagle got all theorists to send us predictions as a function of their main single parameter that characterizes the medium in order to do precision fits to the latest PHENIX  $\pi^0$ data including the correct treatment of correlated experimental systematic errors (Fig.~\ref{fig:pi0pqm} )~\cite{ppg079}. 
Systematic uncertainties of the theory predictions were not considered. 

   The large value of the transport coefficient $\mean{\hat{q}=\mu^2/\lambda}=13.2^{+2.1}_{- 3.2}$~GeV$^2$/fm from the best fit to the PQM model~\cite{PQM}  (where $\mu$ is the average 4-momentum transfer to the medium per mean free path $\lambda$) is a subject of some debate in both the more fundamental QCD community~\cite{BS06} and the more phenomenological community~\cite{fragility}. For instance it was stated in Ref.~\cite{fragility} that ``the dependence of $R_{AA}$ on $\hat{q}$ becomes weaker as $\hat{q}$ increases'' as is clear from Fig.~\ref{fig:pi0pqm}b. It was also asserted that ``when the values of the time-averaged transport coefficient $\hat{q}$ exceeds 5~GeV$^2$/fm, $R_{AA}$ gradually loses its sensitivity.'' That statement also appeared reasonable. However, given the opportunity of looking at a whole range of theoretical predictions (kindly provided by the PQM authors~\cite{PQM}) rather than just the one that happens to fit the data, we experimentalists learned something about the theory that was different from what the theorists emphasized. By simply looking at the PQM predictions on a log-log plot (Fig.~\ref{fig:pi0pqm}c), it became evident that the PQM prediction could be parameterized as $R_{AA}[p_T=20 {\rm~GeV/c}]=0.75/\sqrt{\hat{q}\,({\rm~GeV^2/fm})}$ over the range $5<\hat{q}<100$~GeV$^2$/fm. This means that in this range, the fractional sensitivity to $\hat{q}$ is simply proportional to the fractional uncertainty in $R_{AA}$, i.e.  $\Delta\hat{q}/\hat{q}=2.0\times \Delta R_{AA}/R_{AA}$, so that improving the precision of $R_{AA}$ e.g. in the range $10\leq p_T\leq 20$~GeV/c will lead to improved precision on $\mean{\hat{q}}$. This is a strong incentive for experimentalists. Similarly, this should give the theorists incentive to improve their (generally unstated) systematic uncertainties.  
\subsection{$R_{AA}$ vs. the reaction plane}  
   Another good synergy between experimentalists and theorists is the study of $R_{AA}$ as a function of angle to the reaction plane and centrality in order to understand the effect of varying the initial conditions (centrality) and the path length through the medium (angle). When PHENIX first presented results on $R_{AA}(p_T)$ vs. the angle $\Delta\phi$ to the reaction plane~\cite{ppg054} there was a reaction from the flow community that this is nothing other than a different way to present the anisotropic flow, $v_2$. This is strictly not true for two reasons: 1) $v_2$ measurements are relative while $R_{AA}(\Delta\phi, p_T)$ is an absolute measurement including efficiency, acceptance and all other such corrections; 2) if and only if the angular distribution of high $p_T$ suppression around the reaction plane were simply a second harmonic so that all the harmonics other than $v_2$ vanish (and why should that be?) then $R_{AA}(\Delta\phi, p_T)/R_{AA}(p_T)=1+2 v_2\cos 2\Delta\phi$. Nevertheless, whatever the actual form of the angular distribution, it is true that $R_{AA}(\Delta\phi, p_T)/R_{AA}(p_T)=dN(\Delta\phi, p_T)/d\Delta\phi /\mean{dN(\Delta\phi, p_T)/d\Delta\phi)}$ but without the absolute values it is impossible to tell whether $R_{AA}(\Delta\phi, p_T)$ approaches or exceeds 1 (or any other value) at some value of $\Delta\phi$. 
   
\begin{figure}[!h]
\begin{center}
\includegraphics[width=0.75\linewidth]{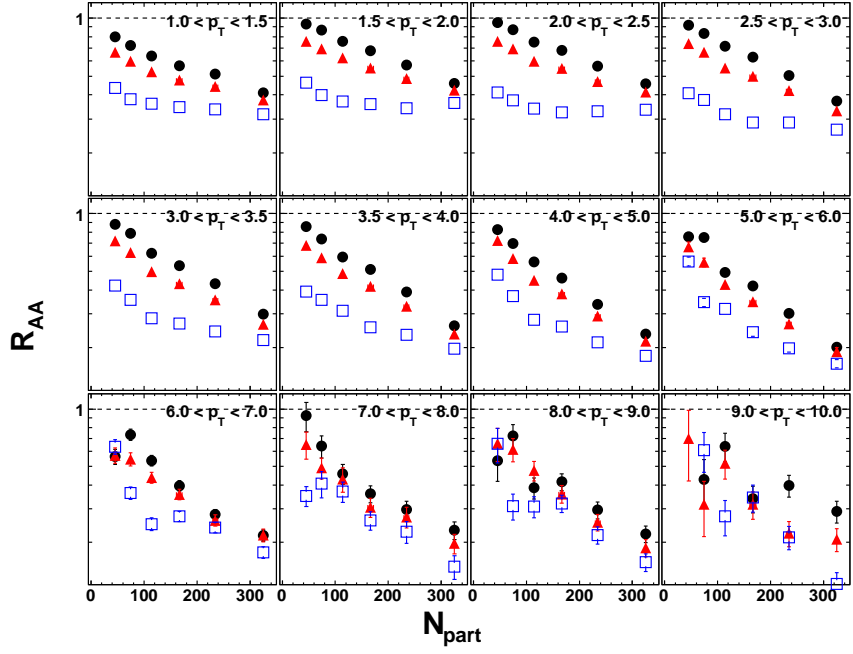} 
\end{center}
\caption[]{Nuclear Modification Factor, $R_{AA}^{\pi^0}$ in reaction-plane bins as a function of $p_T$ and centrality ($N_{\rm part}$) in Au+Au collisions at $\sqrt{s_{NN}}=200$~GeV~\cite{ppg092}. Filled circles represent $R_{AA}(0< \Delta\phi < 15^\circ)$ (in-plane), open squares $R_{AA}(75< \Delta\phi < 90^\circ)$ (out-of-plane) and filled triangles  $R_{AA}(30< \Delta\phi < 45^\circ)$.  
\label{fig:ppg092} }
\end{figure}

   For instance in a new result this year, PHENIX has observed a striking difference in the behavior of the dependence of the in-plane $R_{AA}(\Delta\phi\sim0, p_T)$ for $\pi^0$ as a function of centrality $N_{\rm part}$ compared to the dependence of the $R_{AA}(\Delta\phi\sim \pi/2, p_T)$ in the direction perpendicular to the reaction plane~\cite{ppg092} (Fig.~\ref{fig:ppg092}).  
   The $R_{AA}$ perpendicular to the reaction plane is relatively constant with centrality, while the predominant variation of $R_{AA}$ with centrality comes in the direction parallel to the reaction plane which shows a strong centrality dependence. This is a clear demonstration of the sensitivity of $R_{AA}$ to the length traversed in the medium which is relatively constant as a function of centrality perpendicular to the reaction plane but depends strongly on the centrality parallel to the reaction plane. This is a fantastic but reasonable result and suggests that tests of the many models of energy loss should concentrate on comparing the centrality dependence for directions parallel to the reaction plane, where the length traversed depends strongly on centrality, compared to perpendicular to the reaction plane, where the length doesn't change much with centrality, before tackling the entire angular distribution.    
   
\begin{figure}[!h]
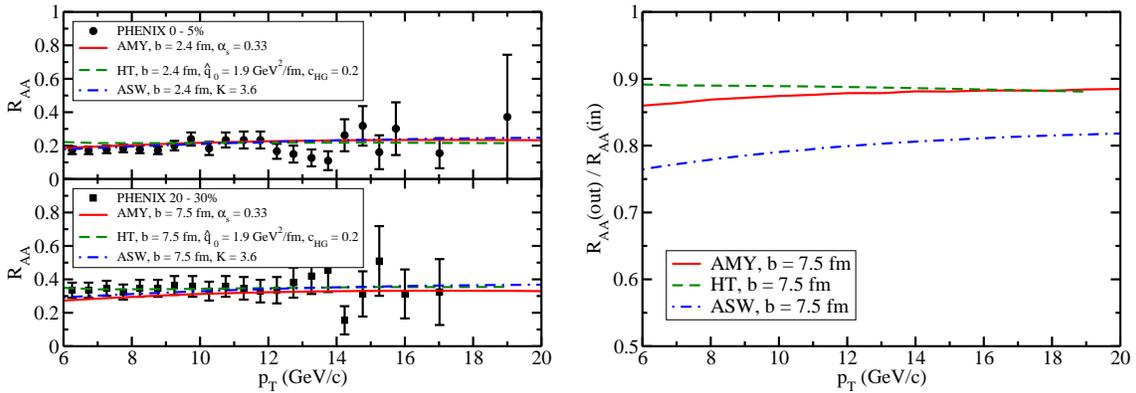
 
\begin{center}
\begin{tabular}{cc}
\hspace*{-0.002\linewidth}\includegraphics[width=0.48\linewidth]{figs/RAA_centrality.eps} &
\hspace*{0.001\linewidth}\includegraphics[width=0.48\linewidth]{figs/RAA_out_in_ratio.eps}
\end{tabular}
\end{center}
\caption[]
{a) (left) Three theoretical models~\cite{Bass*} compared to PHENIX $R_{AA}^{\pi^0}(p_T)$ in Au+Au collisions at 0-5\% and 20-30\% centrality~\cite{pi0-QM05}. b) (right) $R_{AA}$(out)/$R_{AA}$(in) ratio for same 3 models at 20-30\% centrality ($N_{\rm part}=167\pm 5$) .       
\label{fig:th-angle} }
\end{figure}

   The theorists have not been idle on this issue and are making great strides by attempting to put all the theoretical models of jet quenching into a common nuclear geometrical and medium evolution formalism so as to get an idea of the fundamental differences in the models~\cite{Bass*} ``evaluated on identical media, initial state and final fragmentation. The only difference in models will be in the Eloss kernel.''. The different models~\cite{Bass*} all agreed with the measured $R_{AA}(p_T)$ (Fig~\ref{fig:th-angle}a); but the agreement with the measured $R_{AA}(\Delta\phi, p_T)$ as shown by the $R_{AA}$(out)/$R_{AA}$(in) ratio is not very good (Fig~\ref{fig:th-angle}b). Hopefully the latest PHENIX results~\cite{ppg092} (Fig.~\ref{fig:ppg092}) will suggest the way for further  improvement.  

\section{Do direct-$e^{\pm}$ from Heavy Flavors indicate one or two theoretical crises?}
\label{sec:bc}
\begin{figure}[!ht]
\begin{center} 
\begin{tabular}{cc}
\hspace*{-0.04\linewidth}\includegraphics*[width=0.53\linewidth]{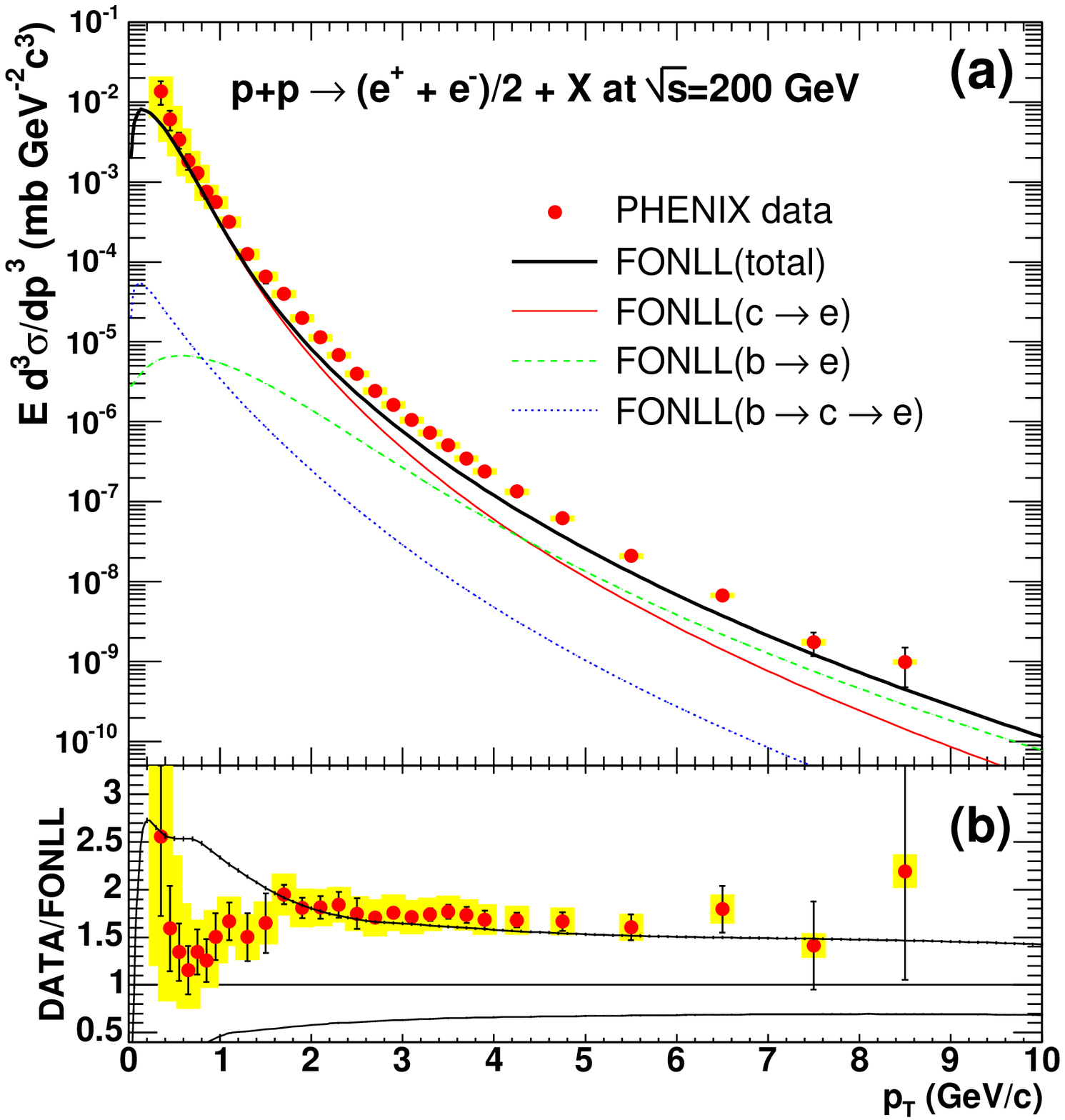} & 
\hspace*{-0.02\linewidth}\includegraphics*[width=0.51\linewidth]{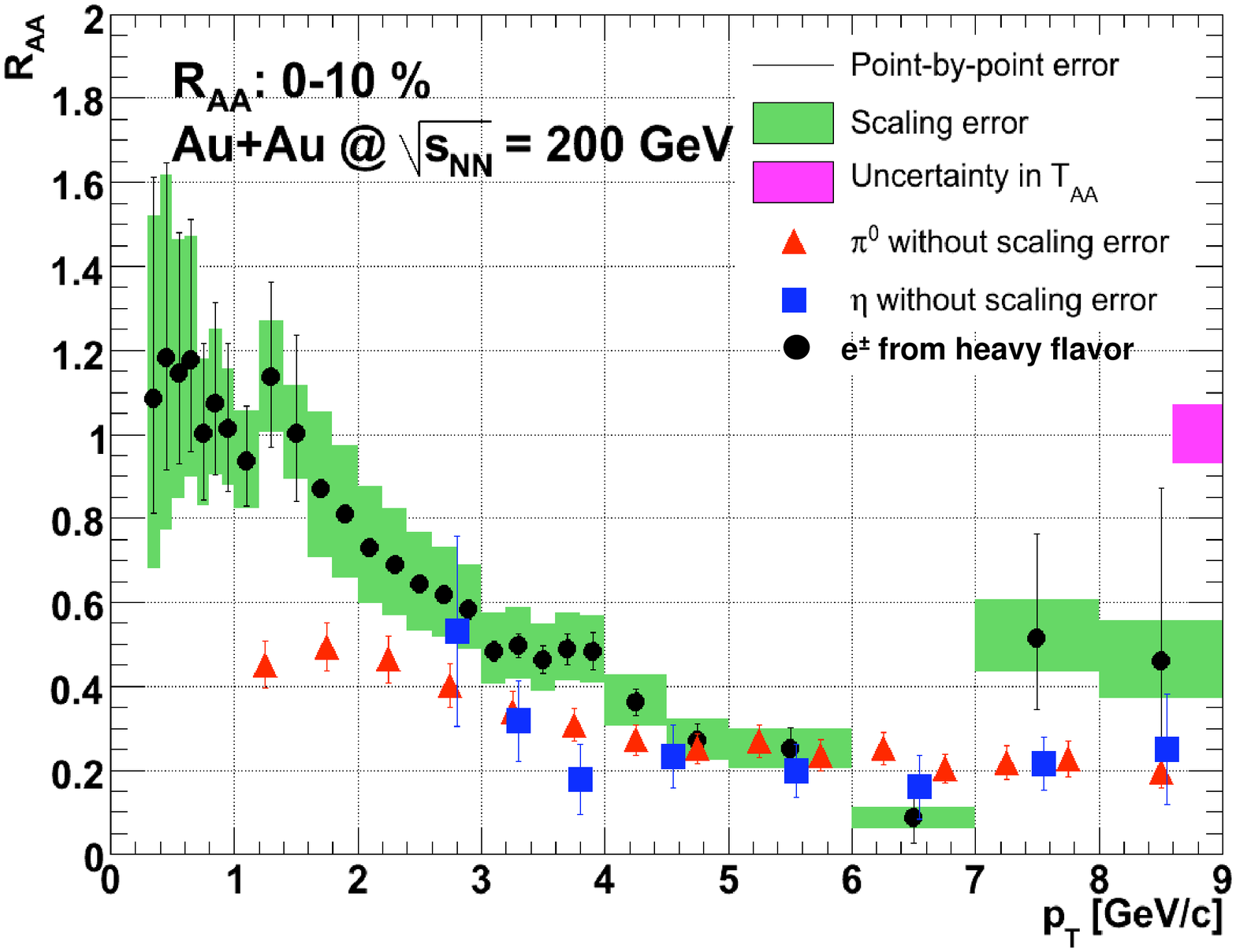} 
\end{tabular}
\end{center}\vspace*{-1.5pc}
\caption[]{a) (left) Invariant cross section of direct single-$e^{\pm}$ in p-p collisions ~\cite{PXcharmpp06} compared to theoretical predictions from $c$ and $b$ quark semileptonic decay.~\cite{forRamona} b) (right) $R_{AA}$ as a function of $p_T$ for direct single-$e^{\pm}$~\cite{PXPRL97e}, $\pi^0$ and $\eta$ in Au+Au central (0-10\%) collisions at $\sqrt{s_{NN}}=200$~GeV.}
\label{fig:f7}
\end{figure}

   PHENIX was specifically designed to be able to detect charm particles via direct single-$e^{\pm}$ from their semileptonic decay. Fig.~\ref{fig:f7}a shows our direct single-$e^{\pm}$ measurement in p-p collisions at $\sqrt{s}=200$~GeV~\cite{PXcharmpp06} in agreement with a QCD calculation~\cite{forRamona} of $c$ and $b$ quarks as the source of the direct single-$e^{\pm}$ (heavy-flavor $e^{\pm}$).   
In Au+Au collisions, a totally unexpected result was observed. The direct single-$e^{\pm}$ from heavy quarks are suppressed the same as the $\pi^0$ and $\eta$ from light quarks (and gluons) in the range $4\leq p_T\leq 9$~GeV/c where $b$ and $c$ contributions are roughly equal (Fig.~\ref{fig:f7}b)~\cite{PXPRL97e}. This strongly disfavors the QCD energy-loss explanation of jet-quenching because, naively, heavy quarks should radiate much less than light quarks and gluons in the medium; but opens up a whole range of new possibilities including string theory~\cite{egsee066}. 
\subsection{Zichichi to the rescue?}
  In September 2007, I read an article by Nino Zichichi, ``Yukawa's gold mine'', in the CERN Courier, taken from his talk at the 2007 International Nuclear Physics meeting in Tokyo, Japan, in which he proposed: ``the reason why the top quark appears to be so heavy (around 200~GeV) could be the result of some, so far unknown, condition related to the fact that the final
state must be QCD-colourless. We know that confinement produces masses of the order of a giga-electron-volt. Therefore, according to our present understanding, the QCD colourless condition cannot explain the heavy quark mass. However, since the origin of the quark masses is still not known, it cannot be excluded that in a QCD coloured world, the six quarks are all nearly massless and that the colourless condition is `flavour' dependent.'' 
  
  Nino's idea really excited me, even though, or perhaps, because, it appeared to overturn two of the major tenets of the Standard Model since it seemed to imply that: QCD isn't flavor blind;  the masses of quarks aren't given by the Higgs mechanism.  Massless $b$ and $c$ quarks in a color-charged medium would be the simplest way to explain the apparent equality of gluon, light quark and heavy quark suppression indicated by the equality of $R_{AA}$ for $\pi^0$ and $R_{AA}$ of direct single-$e^{\pm}$ in regions where both $c$ and $b$ quarks dominate. Furthermore RHIC and LHC-Ions are the only place in the Universe to test this idea. 
  Nino's idea seems much more reasonable to me than the string theory explanations of heavy-quark suppression (especially since they can't explain light-quark suppression). Nevertheless, just to be safe, I asked some distinguished theorists what they thought, with these results:
  ``Oh, you mean the Higgs field can't penetrate the QGP'' (Stan Brodsky);
``You mean that the propagation of heavy and light quarks through the medium is the same'' (Rob Pisarski); 
``The Higgs coupling to vector bosons $\gamma$, $W$, $Z$ is specified in the standard model and is a fundamental issue. One big question to be answered by the LHC is whether the Higgs gives mass to fermions or only to gauge bosons. The Yukawa couplings to fermions are put in by hand and are not required'' ``What sets fermion masses, mixings?" (Chris Quigg-Moriond2008); ``No change in the $t$-quark, $W$, Higgs mass relationship   if there is no Yukawa coupling: but there could be other changes'' (Bill Marciano).

	 Nino proposed to test his idea by shooting a proton beam through a QGP formed in a Pb+Pb collision at the LHC and seeing the proton `dissolved' by the QGP. My idea is to use the new PHENIX vertex detector, to be installed in 2010, to  map out, on an event-by-event basis, the di-hadron correlations from identified $b,\overline{b}$ di-jets, identified $c,\bar{c}$ di-jets, which do not originate from the vertex, and light quark and gluon di-jets, which originate from the vertex and can be measured with $\pi^0$-hadron correlations. These measurements will confirm in detail (or falsify) whether the different flavors of quarks behave as if they have the same mass in a color-charged medium. Depending when the LHC-Ions starts, it is conceivable that ALICE or another LHC experiment with a good vertex detector could beat RHIC to the punch, since this measurement compares the energy loss of light and heavy quarks and may not need p-p comparison data.
	 
	 If Nino's proposed effect is true, that the masses of fermions are not given by the Higgs, and we can confirm the effect at RHIC or LHC-Ions, this would be a case where Relativistic Heavy Ion Physics may have something unique to contribute at the most fundamental level to the Standard Model---a ``transformational discovery.'' Of course the LHC or Tevatron could falsify this idea by finding the Higgs decay to $b,\overline{b}$ at the expected rate in p-p collisions. 

\section{Soft physics projections for LHC}
\begin{figure}[!h] 
\begin{center}
\begin{tabular}{cc}
\hspace*{-0.025\linewidth}\includegraphics[width=0.48\linewidth]{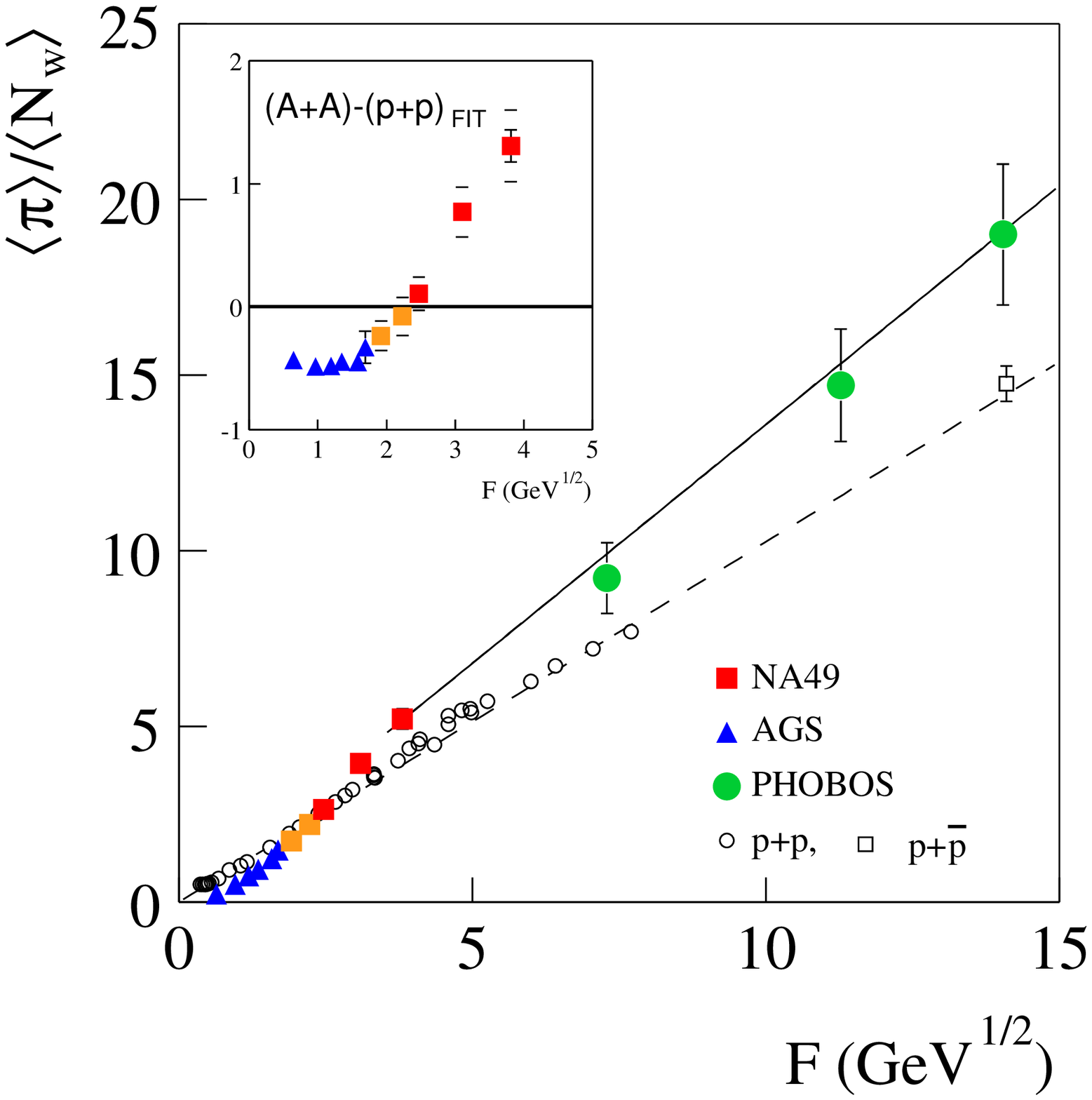} &
\hspace*{-0.02\linewidth}\includegraphics[width=0.52\linewidth]{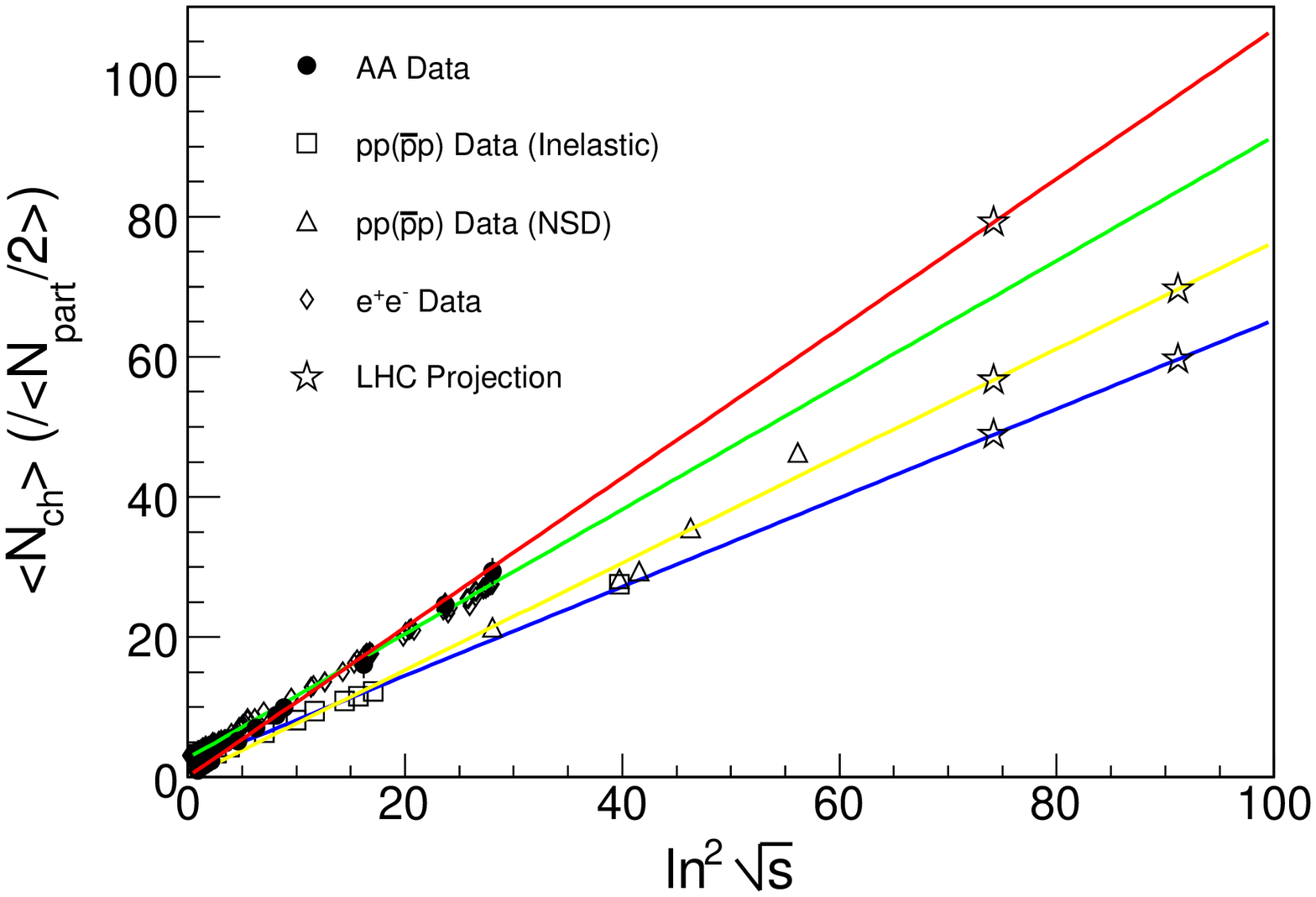}
\end{tabular}
\end{center}
\caption[]
{a) (left) Gazdzciki's plot~\cite{MGQM04} of pions/participant vs. $F=(\sqrt{s_{NN}}-2m_N)^{3/4}/\sqrt{s_{NN}}^{1/4} \approx \sqrt{s_{NN}}^{1/2}$ in A+A and p+p collisions. b) (right) Wit Busza's prediction for the number of charged particles per participant pair vs. $\ln^2 \sqrt{s}$(GeV)~\cite{LastCall}.       
\label{fig:kink} }
\end{figure}

    Some soft physics issues at LHC are also very interesting to me. Marek Gazdzicki has popularized 3 features from the NA49 results~\cite{MGQM04} at the CERN SpS fixed target heavy ion program, which he calls `the kink', `the horn' and `the step'. I believe that `the kink' is certainly correct (Fig.~\ref{fig:kink}a) and has relevance to the LHC program.  The `kink' reflects the fact that the wounded nucleon model (WNM)~\cite{WNM} works only at $\sqrt{s_{NN}}\sim 20$~GeV where it was discovered~\cite{Busza,WA80} and fails above and below $\sqrt{s_{NN}}\sim 20$~GeV: wounded projectile nucleons below 20~GeV at mid-rapidity~\cite{E802-87}; wounded projectile quarks (AQM)~\cite{AQM}, 31~GeV and above~\cite{Ochiai,Nouicer}.
This led me to speculate that maybe the charged-particle multiplicity or sum-transverse energy might be the only quantity to exhibit point-like $N_{\rm coll}$ scaling at LHC energies. However, Wit Busza's prediction for the charged multiplicity per participant pair increased by the same ratio from RHIC to LHC in both A+A and p-p collisions, which implies that the AQM will still work at the LHC. This makes me think that $N_{\rm coll}$ scaling for soft-processes at LHC is unlikely.  

\begin{figure}[!h] 
\begin{center}
\begin{tabular}{ccc}
\hspace*{-0.014\linewidth}\includegraphics[width=0.35\linewidth]{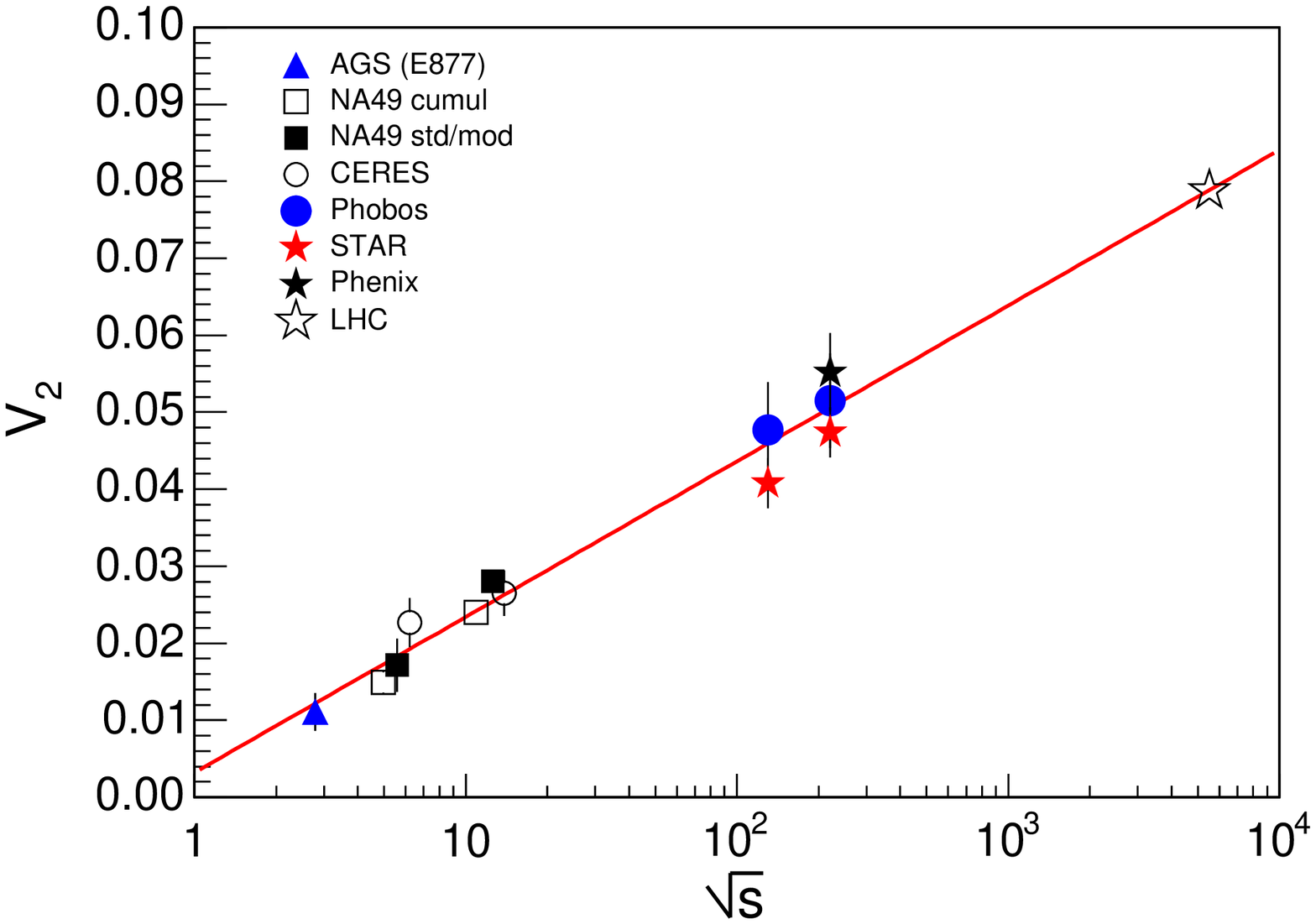} &
\hspace*{-0.027\linewidth}\includegraphics[width=0.33\linewidth]{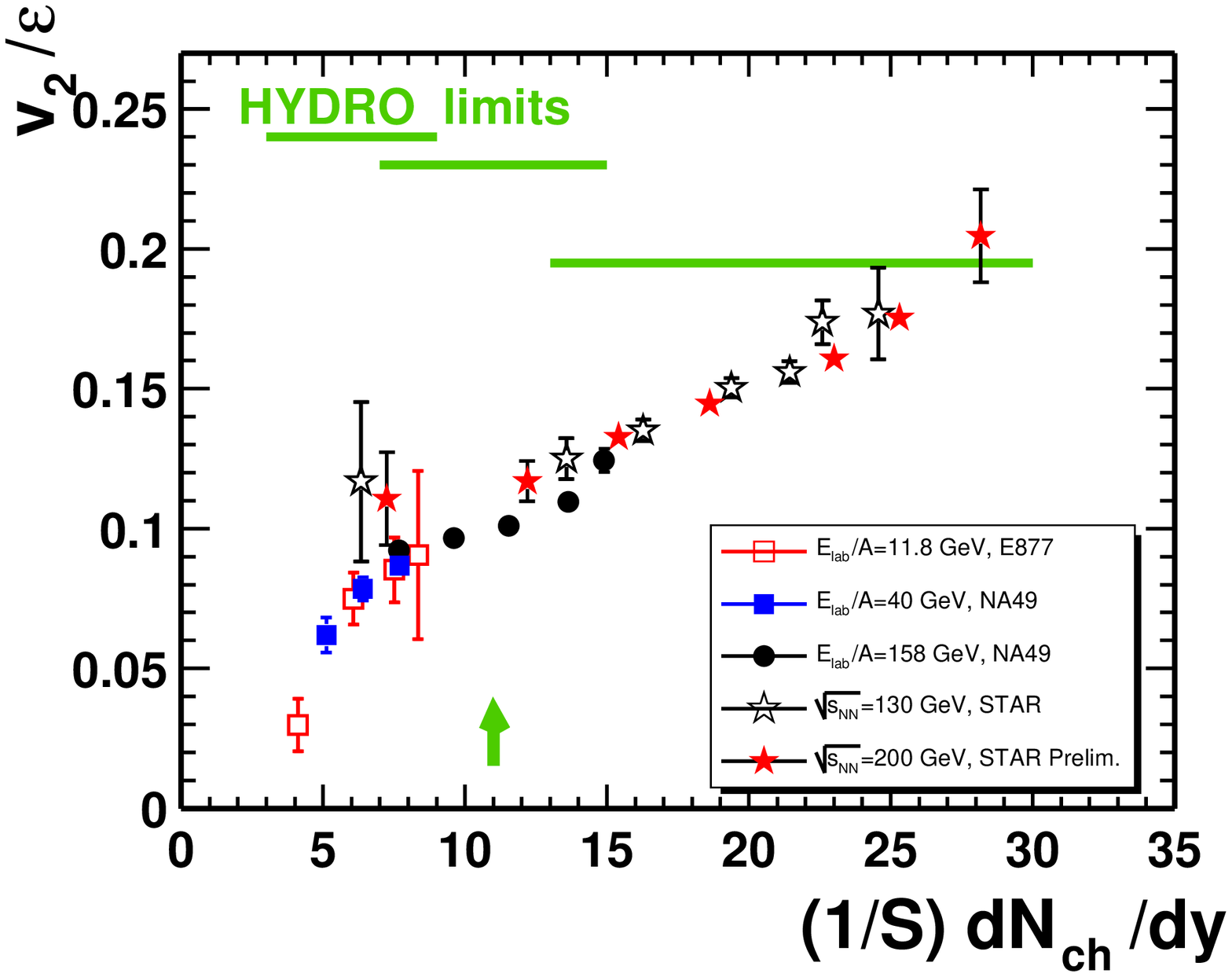}
\hspace*{-0.022\linewidth}\includegraphics[width=0.33\linewidth]{figs/SongHeinzFig7b.eps}
\end{tabular}
\end{center}
\caption[]
{a) (left) Busza's extrapolation of $v_2$ to LHC~\cite{LastCall}. b) (center) $v_2/\varepsilon$ vs `Bjorken multiplicity density', $(1/S) dN_{\rm ch}/dy$~\cite{Alt}. 
c) (right) `Hydro Limit' calculated in viscous Hydrodynamics for several values of the initial energy density $e_0$~\cite{SongHeinz}.      
\label{fig:v2limit} }
\end{figure}

   A more interesting soft physics issue for the LHC concerns the possible increase of the an-isotropic flow $v_2$ beyond the `hydrodynamic limit'. Wit Busza's extrapolation~\cite{LastCall} of $v_2$ to the LHC energy is shown in Fig.~\ref{fig:v2limit}a, a factor of 1.6 increase from RHIC.  
A previous paper by NA49~\cite{Alt} which compared $v_2$ measurements from AGS and CERN fixed target experiments to RHIC as a function of the `Bjorken multiplicity density', $dn_{\rm ch}/d\eta /S$, where $S=$ is the overlap area of the collision zone, showed an increase in $v_2/\varepsilon$ from fixed target energies to RHIC leading to a ``hydro limit'', where $\varepsilon$ is the eccentricity of the collision zone (Fig.~\ref{fig:v2limit}b). This limit was confirmed in a recent calculation using viscous relativistic hydrodynamics~\cite{SongHeinz} which showed a clear hydro-limit of $v_2/\varepsilon=0.20$ (Fig.~\ref{fig:v2limit}c). This limit is sensitive to the ratio of the viscosity/entropy density, the now famous $\eta/s$, but negligibly sensitive to the maximum energy density of the collision, so I assume that this calculation would give a hydro-limit at the LHC not too different from RHIC, $v_2/\varepsilon\approx 0.20$. Busza's extrapolation of a factor of 1.6 increase in $v_2$ from RHIC to LHC combined with $v_2/\varepsilon$ from Fig.~\ref{fig:v2limit}b gives $v_2/\varepsilon=0.32$ at LHC. In my opinion this is a measurement which can be done to high precision on the first day of Pb+Pb collisions at the LHC, since it is high rate and needs no p-p comparison data. Personally, I wonder what the hydro aficionados would say if both Heinz and Busza's predictions were correct?


\begin{thebibliography}{99}
\bibitem{ThanksAM} Thanks to Sasha Milov for the plot of $R_{AA}(p_T)$ for all PHENIX published and preliminary measurements. With the exception of the internal-conversion direct-$\gamma$ where the fit to the p-p data is used to compute $R_{AA}$, all the other values of $R_{AA}$ are computed from Eq.~\ref{eq:RAA} using the measured Au+Au and p-p data points.
\end{thebibliography}
\end{document}